\pdfoutput=1

\documentclass[11pt,twoside,a4paper,cmspaper,final,collab]{cms-tdr}
\def\svnVersion{452715P}\def\cmsCernNoTag{CERN-EP-2017-233}\def\cmsCernDate{\today}\def\cmsMessage{Published in Physics Letters B as \href{http://dx.doi.org/10.1016/j.physletb.2018.02.050}{\doi{10.1016/j.physletb.2018.02.050.}}}

\begin{document}\cmsNoteHeader{HIG-16-044}

\hyphenation{had-ron-i-za-tion}
\hyphenation{cal-or-i-me-ter}
\hyphenation{de-vices}

\RCS$Revision: 452554 $
\RCS$HeadURL: svn+ssh://svn.cern.ch/reps/tdr2/papers/HIG-16-044/trunk/HIG-16-044.tex $
\RCS$Id: HIG-16-044.tex 452554 2018-03-25 21:50:16Z jacobo $

\providecommand{\NA}{\ensuremath{\text{---}}}
\newcommand{\CMVAL}{\ensuremath{\mathrm{CMVA_{L}}}}
\newcommand{\CMVAmax}{\ensuremath{\mathrm{CMVA_{max}}}}
\newcommand{\CMVAmin}{\ensuremath{\mathrm{CMVA_{min}}}}
\newcommand{\CMVAM}{\ensuremath{\mathrm{CMVA_{M}}}}
\newcommand{\CMVAT}{\ensuremath{\mathrm{CMVA_{T}}}}
\newcommand{\dEtaJJ}{\ensuremath{\Delta \eta\mathrm{(jj)}}}
\newcommand{\dPhiJJ}{\ensuremath{\Delta\phi(\mathrm{jj})}}
\newcommand{\dphiVH}{\ensuremath{\Delta\phi(\mathrm{V,jj})}}
\newcommand{\dRJJ}{\ensuremath{\Delta R(\mathrm{jj})}}
\newcommand{\HBB}{\ensuremath{\PH\to\bbbar}\xspace}
\newcommand{\METsig}{\ensuremath{\sigma(\MPT)}}
\newcommand{\MHT}{\ensuremath{H_{\mathrm{T}}^{\text{miss}}}}
\newcommand{\mH}{\ensuremath{m_\PH}\xspace}
\newcommand{\Mjj}{\ensuremath{M\mathrm{(jj)}}}
\newcommand{\Mll}{\ensuremath{M(\ell\ell)}}
\newcommand{\MPT}{\ensuremath{{p}_{\mathrm{T}}^{\text{miss}}}}
\newcommand{\MPTvec}{\ensuremath{\ptvec^{\text{miss}}}}
\newcommand{\MPTtrk}{\ensuremath{\MPT({\text{trk}})}}
\newcommand{\MPTvectrk}{\ensuremath{\MPTvec({\text{trk}})}}
\newcommand{\dphiMtrk}{\ensuremath{\Delta\phi(\MPTvec,\MPTvectrk)}}
\newcommand{\mtop}{\ensuremath{m_{\mathrm{top}}}}
\newcommand{\Naj}{\ensuremath{N_{\mathrm{aj}}}}
\newcommand{\Nal}{\ensuremath{N_{\mathrm{a}\ell}}}
\newcommand{\one}{1-}
\newcommand{\ptjj}{\ensuremath{{\pt\mathrm{(jj)}}}}
\newcommand{\ptll}{\ensuremath{{\pt}(\ell\ell)}}
\newcommand{\ptl}{\ensuremath{p_{\mathrm{T}}^{\ell}}}
\newcommand{\ptone}{\ensuremath{{\pt}({\mathrm{j}}_1})}
\newcommand{\pttwo}{\ensuremath{{\pt}({\mathrm{j}}_2})}
\newcommand{\ptV}{\ensuremath{\pt(\Vvar)}\xspace}
\newcommand{\twol}{2-lepton} \newcommand{\onel}{1-lepton}
\newcommand{\VH}{\ensuremath{{\Vvar}\PH}\xspace}
\newcommand{\Vvar}{\ensuremath{\cmsSymbolFace{V}}\xspace}
\newcommand{\VZ}{\ensuremath{{\Vvar}\cPZ}\xspace}
\newcommand{\WenH}{\ensuremath{\PW(\Pe\cPgn)\PH}\xspace}
\newcommand{\WH}{\ensuremath{\PW\PH}\xspace}
\newcommand{\WlnH}{\ensuremath{\PW(\ell\cPgn)\PH}\xspace}
\newcommand{\Wln}{\ensuremath{\PW(\ell\cPgn)}}
\newcommand{\WmnH}{\ensuremath{\PW(\Pgm\cPgn)\PH}\xspace}
\newcommand{\WtoLN}{\ensuremath{\PW\to\ell\cPgn}}
\newcommand{\WZ}{\ensuremath{\PW\cPZ}\xspace}
\newcommand{\ZBB}{\ensuremath{\cPZ\to\bbbar}\xspace}
\newcommand{\zerol}{0-lepton}
\newcommand{\zero}{0-}
\newcommand{\ZH}{\ensuremath{\cPZ\PH}\xspace}
\newcommand{\ZllHbb}{\ensuremath{\cPZ(\ell\ell)\PH(\cPqb\cPqb)}}
\newcommand{\ZllH}{\ensuremath{\cPZ(\ell\ell)\PH}\xspace}
\newcommand{\ZnnH}{\ensuremath{\cPZ(\cPgn\cPgn)\PH}\xspace}
\newcommand{\Znn}{\ensuremath{\cPZ(\cPgn\cPgn)}}
\newcommand{\ZtoLL}{\ensuremath{\cPZ\to\ell\ell}}
\newcommand{\ZtoNN}{\ensuremath{\cPZ\to\cPgn\cPagn}}
\newlength{\cmsTabSkip}\setlength\cmsTabSkip{1.8ex}
\newlength\cmsFigWidth
\ifthenelse{\boolean{cms@external}}{\setlength\cmsFigWidth{0.98\columnwidth}}{\setlength\cmsFigWidth{0.6\textwidth}}
\newlength\cmsFigWidthVI
\ifthenelse{\boolean{cms@external}}{\setlength\cmsFigWidthVI{0.42\textwidth}}{\setlength\cmsFigWidthVI{0.45\textwidth}}
\ifthenelse{\boolean{cms@external}}{\providecommand{\cmsTabResize}[1]{#1}}{\providecommand{\cmsTabResize}[1]{\resizebox{\textwidth}{!}{#1}}}

\newcommand{\xmark}{\ensuremath{\checkmark}}

\cmsNoteHeader{HIG-15-006}
\title{Evidence for the Higgs boson decay to a bottom quark-antiquark pair}

\date{\today}

\abstract{A search for the standard model (SM) Higgs boson ($\PH$) decaying to $\bbbar$ when produced in association with an electroweak vector boson is reported for the following processes: $\cPZ(\nu\nu)\PH$, $\PW(\mu\nu)\PH$, $\PW(\Pe\nu)\PH$, $\cPZ(\mu\mu)\PH$, and $\cPZ(\Pe\Pe)\PH$. The search is performed in data samples corresponding to an integrated luminosity of 35.9\fbinv at $\sqrt{s}=13\TeV$ recorded by the CMS experiment at the LHC during Run~2 in 2016. An excess of events is observed in data compared to the expectation in the absence of a \HBB signal. The significance of this excess is 3.3 standard deviations, where the expectation from SM Higgs boson production is 2.8. The signal strength corresponding to this excess, relative to that of the SM Higgs boson production, is $1.2 \pm 0.4$. When combined with the Run~1 measurement of the same processes, the signal significance is 3.8 standard deviations with 3.8 expected. The corresponding signal strength, relative to that of the SM Higgs boson, is $1.06^{+0.31}_{-0.29}$.}

\hypersetup{%
pdfauthor={CMS Collaboration},%
pdftitle={Evidence for the Higgs boson decay to a bottom quark-antiquark pair},%
pdfsubject={CMS},%
pdfkeywords={CMS, physics, Higgs}}

\maketitle

\section{Introduction}\label{sec:hbb_Introduction}

The ATLAS and CMS Collaborations reported in 2012 the discovery of a new
boson with a mass near 125\GeV using data from the Large Hadron
Collider (LHC) at CERN~\cite{Aad:2012tfa,Chatrchyan:2012ufa,Chatrchyan:2013lba}. Significant signals have been observed in channels where the boson decays into $\gamma \gamma$,
$\cPZ\cPZ$, $\PW\PW$, or $\Pgt\Pgt$~\cite{Aad:2014eha,Khachatryan:1728107,Aad:2014eva,Chatrchyan:2013mxa,ATLAS:2014aga,Aad:2015ona,Chatrchyan:1633401,Aad:2015vsa,Chatrchyan:1643937,Sirunyan:2273798}.
The measured production and decay rates and spin-parity
properties of this boson~\cite{Aad:2015gba,Khachatryan:2014jba,Chatrchyan:2012jja,Aad:2013xqa,Khachatryan:2016vau,Aad:2015zhl,Sirunyan:2272260} are compatible with those of the standard model (SM) Higgs
boson (H)~\cite{Englert:1964et,Higgs:1964ia,Higgs:1964pj,Guralnik:1964eu,Higgs:1966ev,Kibble:1967sv}.

The \HBB decay tests directly the Higgs boson coupling to
fermions, and more specifically to down-type quarks, and has not yet been established
experimentally. In the SM, for a Higgs boson mass $\mH = 125$\GeV, the branching fraction is approximately 58\%~\cite{Heinemeyer:2013tqa}, by far the largest. An observation in this channel is necessary to solidify the Higgs boson as the source of mass generation in the fermion sector of the SM~\cite{PhysRevLett.19.1264,Nambu:1961fr}.

At the Tevatron \Pp\Pap\ collider the sensitivity of the SM Higgs boson search, for masses below 130\GeV,
was dominated by its production in association with a weak vector boson (\VH\ production) and its decay to \bbbar~\cite{PhysRevLett.109.071804}. The combined searches from the CDF and D0 Collaborations resulted in an excess of events with a local significance, at $\mH=125$\GeV, of 2.8 standard deviations, with an expected value of 1.6. For the \HBB search at the LHC, the following Higgs boson production processes have been considered: in association with a top quark pair~\cite{Aad:2015gra,Aad:2016zqi,Chatrchyan:2013yea,Khachatryan:2014qaa}, through
vector boson fusion~\cite{Aaboud:2016cns,Khachatryan:2015bnx}, through \VH\ production~\cite{Aad:2014xzb,Chatrchyan:1610290}, and, more recently, through gluon fusion~\cite{CMS-PAS-HIG-17-010}. The process with the largest sensitivity is \VH\ production.

The combined  searches for \HBB  by the ATLAS and CMS Collaborations in Run~1, at $\sqrt{s} = 7$ and 8\TeV, evaluated for a Higgs boson mass of 125.09\GeV, resulted in a significance of 2.6 standard deviations, with 3.7 standard deviations expected~\cite{Khachatryan:2016vau}. The corresponding signal strength, relative to the SM expectation, is $\mu = 0.7 \pm 0.3$.
The significance from the individual search by the ATLAS (CMS) experiment is 1.7 (2.0)  standard deviations, with 2.7 (2.5)  standard deviations expected, and a signal strength $\mu = 0.6 \pm 0.4$ ($\mu = 0.8 \pm 0.4$).

Recent results by the ATLAS Collaboration~\cite{Aaboud:2017xsd} in the search for \HBB through VH production at $\sqrt{s} = 13$\TeV, with data corresponding to an integrated luminosity of 36.1\fbinv, report
a significance of 3.5 standard deviations, corresponding to a signal strength of $\mu = 1.20^{+0.42}_{-0.36}$.  The combination with the results from the same search in Run~1~\cite{Aad:2014xzb} yields a significance of 3.6 standard deviations and a signal strength  $\mu = 0.90^{+0.28}_{-0.26}$.

This article reports on the search with the CMS experiment for the decay of the SM Higgs boson to bottom quarks, $\HBB$,
when produced through the $\Pp\Pp\to \VH$ process, where \Vvar is either a \PW\ or a \cPZ\ boson.
This search is performed with data samples from Run~2 of the LHC, recorded during 2016, corresponding to an integrated luminosity of 35.9\fbinv at $\sqrt{s}=13\TeV$. The following five processes are
considered in the search: $\cPZ(\nu\nu)\PH$, $\PW(\mu\nu)\PH$, $\PW(\Pe\nu)\PH$,
$\cPZ(\mu\mu)\PH$, and $\cPZ(\Pe\Pe)\PH$. The final states that
predominantly correspond to these processes, respectively, are characterized by the number of leptons required in the event selection, and are referred to as the \zero, \one, and \twol\ channels.

Throughout this article the term ``lepton'' (denoted $\ell$) refers solely to muons and electrons,
but not to taus. The leptonic tau decays in \WH\ and \ZH\ processes are implicitly included in the
 $\PW(\mu\nu)\PH$, $\PW(\Pe\nu)\PH$, $\cPZ(\mu\mu)\PH$, and $\cPZ(\Pe\Pe)\PH$ processes.
Background processes originate from the production of \PW\ and \cPZ\  bosons in association with
jets from gluons and from light- or heavy-flavor quarks ({\PW}+jets and {\cPZ}+jets), from singly and pair-produced top quarks (single top and \ttbar), from diboson production ({\Vvar}{\Vvar}), and from quantum chromodynamics multijet events (QCD).

Simulated samples of signal and background events are used to optimize the search.
For each channel, a signal region enriched in \VH\ events is selected together with
several control regions, each enriched in events from individual background processes. The control regions are used to test the accuracy
of the simulated samples' modeling for the variables relevant to the analysis. A simultaneous
binned-likelihood fit to the shape and normalization of specific distributions for the signal and
control regions for all channels combined is used to extract a possible Higgs
boson signal.
The distribution used in the signal region is the output of a
boosted decision tree (BDT) event discriminant~\cite{Roe:2004na,Hocker:2007ht}
that helps separate signal from background.  For the control regions,
a variable that identifies jets originating from b quarks, and that discriminates between the different background processes, is used.
To validate the analysis procedure, the same methodology is used to extract a signal
for the \VZ\ process, with \ZBB, which has a nearly identical final state to \VH\,
with \HBB, but with a production cross section of 5 to 15 times larger,
depending on the kinematic regime considered. Finally, the results from this search are combined with those of similar searches performed by the CMS Collaboration during Run~1~\cite{Chatrchyan:1610290,Khachatryan:2015bnx,Khachatryan:2016vau}.

This article is structured as follows: Sections~\ref{sec:hbb_Simulations}--\ref{sec:hbb_Triggers} describe the CMS detector, the simulated samples used for signal and background processes, and the triggers used to collect the data. Sections~\ref{sec:hbb_Event_Reconstruction}--\ref{sec:hbb_Event_Selection} describe the reconstruction of the detector objects used in the analysis and the selection criteria for events in the signal and control regions. Section~\ref{sec:hbb_Uncertainties} describes the sources of uncertainty in the analysis, and Section~\ref{sec:hbb_Results} describes the results, summarized in Section~\ref{sec:hbb_Conclusions}.

\section{The CMS detector and simulated samples}\label{sec:hbb_Simulations}

A detailed description of the CMS detector can be found
elsewhere in Ref.~\cite{Chatrchyan:2008aa}.
The momenta of charged particles are measured using a silicon pixel
and strip tracker that covers the range
$\abs{\eta} < 2.5$ and is immersed in a 3.8\unit{T}
axial magnetic field. The pseudorapidity is defined as $\eta = -\ln[\tan(\theta/2)]$, where $\theta$
is the polar angle of the trajectory of a particle with respect to
the direction of the counterclockwise proton beam.
Surrounding the tracker are a crystal electromagnetic calorimeter
(ECAL) and a brass and scintillator hadron calorimeter (HCAL), both used to
measure particle energy deposits and both consisting of a barrel assembly and two endcaps. The ECAL
and HCAL extend to a range of $\abs{\eta} < 3.0$. A
steel and quartz-fiber Cherenkov forward detector extends the calorimetric
coverage to $\abs{\eta} < 5.0$. The outermost component of the CMS detector is the
muon system, consisting of gas-ionization detectors placed in the
steel flux-return yoke of the magnet
to measure the momenta of muons traversing through the detector. The two-level CMS trigger system selects events of interest for
permanent storage. The first trigger level,
composed of custom hardware processors, uses information from the
calorimeters and muon detectors to select events in less than 3.2\mus.
The high-level trigger software algorithms, executed on a farm of
commercial processors, further reduce the
event rate using information from all detector subsystems. The
variable $\Delta R {=} \sqrt {\smash[b]{(\Delta\eta)^2 {+} (\Delta\phi)^2}}$ is used to
measure the separation between reconstructed objects in the detector,
where $\phi$ is the angle (in radians) of the trajectory of the object in the
plane transverse to the direction of the proton beams.

Samples of simulated signal and background events are produced using
the Monte Carlo (MC) event generators listed below. The CMS detector response is modeled
with \GEANTfour~\cite{GEANT4}.
The signal samples used have Higgs bosons with $\mH = 125$\GeV produced in association with vector bosons.
The quark-induced ZH and WH processes are generated at next-to-leading order (NLO) using the
{\POWHEG}~\cite{Nason:2004rx,POWHEG,Alioli:2010xd} v2 event generator
extended with the MiNLO procedure~\cite{Hamilton2012,Luisoni:2013kna}, while the gluon-induced ZH processes (denoted ggZH)
are generated at leading-order (LO) accuracy with {\POWHEG} v2.
The {\MGvATNLO}~\cite{Alwall:2014hca} v2.3.3 generator is used
at NLO with the FxFx merging scheme~\cite{Frederix:2012ps} for the diboson background samples.
The same generator is used at LO accuracy with the MLM matching scheme~\cite{Alwall:2007fs}
for the {\PW}+jets and {\cPZ}+jets in inclusive and b-quark enriched configurations, as well as the QCD multijet sample.
The \ttbar~\cite{Frixione:2007nw} production process, as well as
the single top quark sample for the $t$-channel~\cite{Frederix:2012dh}, are produced with {\POWHEG} v2.
The single top quark samples for the \cPqt\PW-~\cite{Re:2010bp} and $s$-channel~\cite{Alioli:2009je} are instead produced with {\POWHEG} v1.
The production cross sections for the signal samples are rescaled to
next-to-next-to-leading order (NNLO) QCD + NLO electroweak accuracy combining the
\textsc{vhnnlo}~\cite{Ferrera:2013yga,Ferrera:2014lca,Ferrera:2011bk},
\textsc{vh@nnlo}~\cite{Brein:2012ne,Harlander:2013mla} and \textsc{hawk} v2.0~\cite{Denner:2014cla} generators
as described in the documentation produced by the LHC Working Group on
Higgs boson cross sections~\cite{deFlorian:2016spz}, and they are applied as a function
of the vector boson transverse momentum (\pt).
The production cross sections for the \ttbar
samples are rescaled to the NNLO with the next-to-next-to-leading-log (NNLL) prediction obtained with {{\textsc{Top++} v2.0}}~\cite{Czakon:2011xx},
while
the {\PW}+jets and {\cPZ}+jets samples are rescaled to the NLO cross sections using {\MGvATNLO}.
The parton distribution functions (PDFs) used to produce the NLO samples are
the NLO {NNPDF3.0} set~\cite{Ball:2014uwa}, while the LO NNPDF3.0 set is used for the LO samples.
For parton showering and hadronization the {\POWHEG} and {\MGvATNLO} samples are interfaced with
{\PYTHIA} 8.212~\cite{Sjostrand:2007gs}.
The {\PYTHIA}8 parameters for the underlying event description correspond to the
CUETP8M1 tune derived in Ref.~\cite{Khachatryan:2015pea} based on the work described
in Ref.~\cite{Skands:2014pea}.

During the 2016 data-taking period the LHC instantaneous luminosity reached approximately
$1.5\times 10^{34}\percms$ and the average number of $\Pp\Pp$ interactions per bunch
crossing was approximately 23.  The simulated samples include these additional $\Pp\Pp$
interactions, referred to as pileup interactions (or pileup), that overlap with the event of
interest in the same bunch crossing.

\section{Triggers}\label{sec:hbb_Triggers}

Several triggers are used to collect events with final-state objects
consistent with the signal processes in the channels under consideration.

For the
\zerol\ channel, the quantities used in the trigger are derived from the reconstructed objects
in the detector identified by a particle-flow (PF) algorithm~\cite{Sirunyan:2017ulk} that combines the
online information from all CMS subsystems to identify and reconstruct individual
particles emerging from the proton-proton collisions: charged hadrons, neutral hadrons,
photons, muons, and electrons. The main trigger used requires that both the missing transverse
momentum, \MPT, and the hadronic missing transverse momentum, \MHT, in the event be above a
threshold of 110\GeV.  Online \MPT\ is defined as the magnitude of the negative vector sum of
the transverse momenta of all reconstructed objects identified by the PF algorithm, while \MHT\
is defined as the magnitude of the negative vector sum of the transverse momenta of all
reconstructed jets (with $\pt>20$\GeV and $\abs{\eta}<5.2)$ identified by the same algorithm. For
\ZnnH\ events with $\MPT> 170$\GeV, evaluated offline, the trigger efficiency is approximately
92\%, and near 100\% above 200\GeV.

For the
\onel\ channels, single-lepton triggers are used.
The muon trigger \pt\ threshold is 24\GeV and the electron \pt\ threshold is 27\GeV. For the
\twol\ channels, dilepton triggers are used. The muon \pt\ thresholds are 17 and 8\GeV, and the electron \pt\ thresholds are 23 and 12\GeV.
All leptons in these triggers are required to pass
stringent lepton identification criteria. In addition, to maintain an acceptable trigger rate, and to be consistent
with what is expected from signal events, leptons are also required to be isolated from other tracks and
calorimeter energy deposits. For \WmnH\ events that pass all offline requirements described
in Section~\ref{sec:hbb_Event_Selection}, the single-muon trigger efficiency is ${\approx}$95\%. The corresponding efficiency for recording \WenH\ events with the single-electron trigger
 is ${\approx}$90\%. For \ZllH\ signal events that pass all offline requirements in Section~\ref{sec:hbb_Event_Selection}, the dilepton triggers are nearly 100\% efficient.

\section{Event reconstruction}\label{sec:hbb_Event_Reconstruction}

The characterization of \VH\ events in the channels studied here
requires the reconstruction of the following objects in the detector, using
the PF algorithm~\cite{Sirunyan:2017ulk} and originating from the primary interaction vertex:
muons, electrons, neutrinos (reconstructed as \MPT), and jets --- including those that originate from the hadronization of \cPqb\ quarks, referred to as ``\cPqb\ jets''.

The reconstructed vertex with the largest value of summed physics-object $\pt^2$ is taken to be the primary $\Pp\Pp$ interaction vertex. The physics objects are the objects reconstructed by a jet finding algorithm~\cite{Cacciari:2008gp,Cacciari:fastjet1} applied to all charged tracks associated with the vertex, plus the corresponding associated \MPT.  The pileup interactions affect jet momentum
reconstruction, \MPT reconstruction,
lepton isolation, and \cPqb\ tagging efficiencies.
To mitigate these effects,  all charged hadrons that do not
originate from the primary interaction vertex are removed from consideration in the event.
In addition, the average neutral energy density  from pileup interactions is
evaluated  from PF objects and
subtracted from the reconstructed jets in the
event and from the summed energy in the isolation criteria used for
leptons~\cite{Cacciari:subtraction}.
These pileup mitigation procedures are applied on an object-by-object basis.

Muons are reconstructed using two algorithms~\cite{Chatrchyan:2012xi}: one in which
tracks in the silicon tracker are matched to hits in the muon
detectors, and another in which a track fit is performed using hits in the silicon
tracker and in the muon systems.  In the latter algorithm, the muon is seeded by hits in the
muon systems.  The muon candidates used in the analysis are required to be successfully
reconstructed by both algorithms.  Further identification criteria  are
imposed on the muon candidates to reduce the fraction
of tracks misidentified as muons. These include the number of hits in the tracker and
in the muon systems, the fit quality of the global muon track, and its consistency with
the primary vertex. Muon candidates are required to be in the  $\abs{\eta} < 2.4$ region.

Electron reconstruction~\cite{Khachatryan:2015hwa} requires the matching of a set of ECAL clusters, denoted supercluster (SC),
to a track in the silicon tracker. Electron identification~\cite{Khachatryan:2015hwa} relies on a
multivariate technique that combines observables sensitive to the
amount of bremsstrahlung along the electron trajectory, such as the
geometrical matching and momentum consistency
between the electron trajectory and
the associated calorimeter clusters, as well as various shower shape observables in the calorimeters. Additional requirements are imposed to remove electrons
that originate from photon conversions. Electrons are
required to be in the range $\abs{\eta} < 2.5$,
excluding candidates for which the SC lies in the $1.444 < \abs{\eta_{\mathrm{SC}}}< 1.566$ transition
region between the ECAL barrel and endcap, where electron
reconstruction is not optimal.

Charged leptons from \PW\ and \cPZ\ boson decays are expected to be isolated
from other activity in the event.  For each lepton candidate, a cone in $\eta\NA\phi$
is constructed around the track direction at the event vertex.
The scalar sum of the transverse momentum of each reconstructed
particle, including neutral particles, compatible with the primary vertex and contained within the cone is calculated,
excluding the contribution from the lepton candidate itself.  This sum is called
isolation.  In the presence of pileup,
isolation is contaminated with particles from the other interactions.  A quantity
proportional to the pileup is used to correct the isolation on average to
mitigate reductions in signal efficiency at larger values of pileup. In the \onel\ channel,
if the corrected isolation sum exceeds 6\% of the lepton candidate \pt, the lepton is rejected.
In the \twol\ channel, the threshold is looser; the isolation of each candidate can be up to
20 (15\%) of the muon (electron) \pt.
Including the isolation requirement, the total efficiency for reconstructing
muons is in the range of 85--100\%, depending on \pt and $\eta$. The
corresponding efficiency for electrons is in the range of 40--90\%.

Jets are reconstructed from PF objects using the
anti-\kt clustering algorithm~\cite{Cacciari:2008gp}, with a distance parameter of 0.4,
as implemented in the \FASTJET
package~\cite{Cacciari:fastjet1,Cacciari:fastjet2}.  Each jet is required to
lie within $\abs{\eta} < 2.4$, to have at least two tracks associated with it,
and to have electromagnetic and hadronic energy fractions of at least
1\%. The last requirement removes jets originating from
instrumental effects. Jet energy corrections are applied as a function of $\eta$ and
\pt of the jet~\cite{Chatrchyan:1369486}.  The
missing transverse momentum vector, \MPTvec, is calculated offline as the negative
of the vectorial sum of transverse momenta of all PF objects identified in the
event~\cite{CMS-PAS-JME-16-004}, and the magnitude of this vector is denoted \MPT\ in the
rest of this article.

The identification of \cPqb\ jets is performed using
a combined multivariate
(CMVA) \cPqb\ tagging
algorithm~\cite{BTV12001,CMS-PAS-BTV-15-001}.
This algorithm combines, in a likelihood discriminant, information within jets that helps differentiate
between \cPqb\ jets and jets originating from light quarks, gluons, or
charm quarks. This information includes track impact parameters, secondary
vertices, and information related to low-\pt leptons if contained within a jet.
The output of this discriminant has continuous values between $-1.0$
and 1.0. A jet with a CMVA discriminant value above a certain threshold is labeled as
``b-tagged''. The efficiency for  tagging \cPqb\ jets and the rate of
misidentification of non-\cPqb\ jets depend on the
threshold chosen, and are typically parameterized as a function of the
\pt and $\eta$ of the jets.
These performance measurements are obtained directly from data in
samples that can be enriched in \cPqb\ jets, such as $\ttbar$ and multijet
events (where, for example, requiring the presence of a muon in the
jets enhances the heavy-flavor content of the events). Three thresholds for the CMVA discriminant value are used in this analysis: loose (\CMVAL), medium (\CMVAM), and tight (\CMVAT).
Depending on the threshold used, the efficiencies for tagging jets
that originate from \cPqb\ quarks, \cPqc\ quarks, and light quarks or gluons are in
the 50--75\%, 5--25\%, and 0.15--3.0\% ranges, respectively.
The loose (tight) threshold has the highest (lowest)
efficiency and allows most (least) contamination.

In background events, particularly \ttbar, there is often additional, low energy, hadronic activity
in the event.  Measuring the hadronic activity associated with the main primary vertex
provides additional discriminating variables to reject background.  To measure this hadronic
activity only reconstructed charged-particle tracks are used, excluding
those associated with the vector boson and the two b jets.
A collection of ``additional tracks'' is assembled using reconstructed tracks that:
(i)  satisfy the high purity quality requirements defined in
Ref.~\cite{Chatrchyan:2014fea} and $\pt>300\MeV$;
(ii) are not associated with the vector boson, nor with the selected b jets in the event;
(iii) have a minimum longitudinal impact parameter, $\abs{d_z(\mathrm{PV})}$, with respect to the main PV,
rather than to other pileup interaction vertices;
(iv) satisfy $\abs{d_z(\mathrm{PV})}<2\unit{mm}$; and
(v) are not in the region between the two selected \PQb-tagged jets.
This region is defined as an ellipse in the $\eta\NA\phi$ plane, centered on the midpoint between the two jets,
with major axis of length $\Delta R({\PQb\PQb}){+}1$, where $\Delta R({\PQb\PQb}){=}\sqrt{(\Delta\eta_{\PQb\PQb})^2 {+} (\Delta\phi_{\PQb\PQb})^2}$,
oriented along the direction connecting the two $\PQb$ jets, and with minor axis of length 1.
The additional tracks are then clustered into ``soft-track jets'' using the anti-\kt clustering algorithm with a distance parameter of 0.4.
The use of track jets represents a clean and validated method~\cite{CMS-PAS-JME-10-006} to reconstruct the hadronization of partons with energies down to a few \GeV~\cite{CMS-PAS-JME-08-001};
an extensive study of the soft-track jet activity
can be found in Refs.~\cite{Chatrchyan:2013jya,Khachatryan:2014dea}.  The number of soft
track jets with $\pt>5$\GeV is used in all channels as a background discriminating variable.

Events from data and from the simulated samples are required to
satisfy the same trigger and event reconstruction requirements. Corrections
that account for the differences in the performance of these
algorithms between data and simulated samples are computed from data and used in
the analysis.

\section{Event selection}\label{sec:hbb_Event_Selection}

A signal region enriched in \VH\ events is
determined separately for each channel. Simulated events in this region are
used to train an event BDT discriminant to help differentiate between signal
and background events.  Also for each channel, different control regions,
each enriched in events from individual background processes, are selected. These
regions are used to study the agreement between simulated samples and data, and to provide
a distribution that is combined with the output distribution of the
signal region event BDT discriminant in the $\HBB$ signal-extraction fit.
This control region distribution is obtained from the second-highest value of the CMVA discriminant among the two jets selected for the reconstruction of the \HBB decay, denoted \CMVAmin.

As mentioned in the Introduction, background processes to \VH\ production with \HBB are the production of
vector bosons in association with one or more jets ({\Vvar}+jets), \ttbar
production, single-top-quark production, diboson production, and QCD multijet
production.  These processes have production cross sections that are several
orders of magnitude larger than that of the Higgs boson, with the exception of
the ${\Vvar}\cPZ$ process with $\cPZ\to\bbbar$, with an inclusive cross section only about 15 times larger
than the \VH\ production cross section.  Given the nearly identical final state,
this process provides a benchmark against which the Higgs boson search strategy
can be tested.  The results of this test are discussed in Section~\ref{sec:diboson}.

Below we describe the selection criteria used to define the signal regions
and the variables used to construct the event BDT discriminant.  Also described are
the criteria used to select appropriate background-specific control regions and
the corresponding distributions used in the signal-extraction fit.

\subsection{Signal regions}
\label{sec:hbb_SR}

The signal region
requirements are listed in Table~\ref{tab:PreSel}. Events are selected to belong exclusively to only one of the three channels.
Signal events are characterized by the presence of a vector boson recoiling
against two \cPqb\ jets with an invariant mass near 125\GeV. The event selection
therefore relies on the reconstruction of the decay
of the Higgs boson into two \cPqb-tagged jets and on the reconstruction of the
leptonic decay modes of the vector boson.

The reconstruction of the \HBB decay is based on the selection of the pair
of jets that have the highest values of the CMVA discriminant
among all jets in the event.
The highest and second-highest values
of the CMVA discriminant for these two jets are denoted by \CMVAmax\ and \CMVAmin, respectively.
Both jets are required to be
central (with $\abs{\eta}<2.4$), to satisfy standard requirements to
remove jets from pileup~\cite{CMS-PAS-JME-16-003}, and to have a \pt above a minimum threshold,
that can be different for the highest ($\mathrm{j}_1$) and second-highest ($\mathrm{j}_2$) \pt\ jet.
The selected dijet pair is denoted by ``jj" in the rest of this article.

The background from {\Vvar}+jets and diboson
production is reduced significantly when the \cPqb\ tagging requirements are applied. As a result, processes where the two jets originate from genuine \cPqb\ quarks dominate the sample composition in the signal region. To provide additional suppression of background events, several other requirements are imposed on each channel after the reconstruction
of the \HBB  decay.

\subsubsection{\zerol\ channel}

This channel targets mainly \ZnnH\ events in which the \MPT\ is interpreted as the
transverse momentum of the Z boson in the \ZtoNN\ decay.  In order to overcome large
QCD multijet backgrounds, a relatively high threshold of $\MPT>170$\GeV is required.
The QCD multijet background is further reduced to negligible levels in this channel
when requiring that the \MPT\ does not originate from the direction of (mismeasured)
jets.  To that end, if there is a jet with $\abs{\eta}<2.5$ and $\pt>30$\GeV,
whose azimuthal angle is within 0.5 radians of the \MPT\ direction, the event is rejected.  The rejection of multijet events with \MPT\ produced by mismeasured jets  is aided by using a different missing
transverse momentum reconstruction, denoted \MPTtrk, obtained by considering only charged-particle tracks with $\pt>0.5\GeV$ and $\abs{\eta}<2.5$. For an event to be accepted, it is required that \MPTtrk\ and \MPT\ be aligned in azimuth within 0.5 radians. To
reduce background events from \ttbar\ and \WZ\ production channels, events with any
additional isolated leptons with $\pt>20\GeV$ are rejected.  The number of these additional
leptons is denoted by \Nal.

\begin{table*}[tbp]
\topcaption{Selection criteria that define the signal region. Entries marked
with ``\NA'' indicate that the variable is not used in the given channel.
Where selections differ for different \ptV regions, there are comma separated
entries of thresholds or square brackets with a range that indicate each region's
selection as defined in the first row of the table.
The values listed for kinematic variables are in units of \GeV,
and for angles in units of radians. Where selection differs between lepton flavors, the
selection is listed as (muon, electron).}
\label{tab:PreSel}
\centering
{
\begin{tabular}{lcccc} \hline
Variable                                  & \zerol\                   & \onel\                   & \twol\                 \\
\hline

\ptV                                      & $>$170                    & $>$100                   & $[50,150]$,\,$>$150 \\
\Mll                                      & \NA                         & \NA                        & $[75,105]$          \\
$\ptl$                                    & \NA                         & $(>25,>30)$              & $>$20               \\
$\ptone$                                  & $>$60                     & $>$25                    & $>$20               \\
$\pttwo$                                  & $>$35                     & $>$25                    & $>$20               \\
\ptjj                                     & $>$120                    & $>$100                   & \NA                  \\
\Mjj                                      & $[60,160]$                & $[90,150]$               & $[90,150]$          \\
\dphiVH                                   & $>$2.0                    & $>$2.5                   & $>$2.5              \\
${\mathrm{CMVA_{max}}}$                   & $>$\CMVAT                 & $>$\CMVAT                & $>$\CMVAL          \\
${\mathrm{CMVA_{min}}}$                   & $>$\CMVAL                & $>$\CMVAL                 & $>$\CMVAL          \\
\Naj                                      & $<$2                      & $<$2                     & \NA                  \\
\Nal                                      & $=$0                      & $=$0                     & \NA                  \\
\MPT                                      & $>$170                    & \NA                        & \NA                  \\
$\Delta\phi(\MPTvec,\mathrm{j})$    & $>$0.5                       & \NA                        & \NA                  \\
$\dphiMtrk$                               & $<$0.5                    & \NA                        & \NA                   \\
$\Delta\phi(\MPTvec,\ell)$                   & \NA                         & $<$2.0                   & \NA                   \\
Lepton isolation                          & \NA                         & $<$0.06                  & $(<0.25,<0.15)$     \\
Event BDT                                 & $>-0.8$                   & $>$0.3                   & $>-0.8$              \\
\hline
\end{tabular}
}
\end{table*}

\subsubsection{\onel\ channel}

This channel targets mainly \WlnH\ events in which candidate \WtoLN\ decays are identified
by the presence of one isolated lepton as well as missing transverse momentum, which is implicitly
required in the \ptV selection criteria mentioned below, where \ptV is calculated from the
vectorial sum of \MPTvec and the lepton \ptvec.  Muons (electrons) are required to have
$\pt>25$ $(30)$\GeV.  It is also required that the azimuthal angle between the \MPT\ direction
and the lepton be less than 2.0 radians. The lepton isolation for either flavor of lepton is required
to be smaller than 6\% of the lepton \pt.  These requirements significantly reduce possible
contamination from QCD multijet production.  With the same motivation as in the \zerol\ channel,
events with any additional isolated leptons are rejected.  To substantially reject \ttbar
events, the number of additional jets with $\abs{\eta}<2.9$ and $\pt>25\GeV$, \Naj, is allowed to be at most one.

\subsubsection{\twol\ channel}

This channel targets \ZtoLL\ decays, which are reconstructed by combining isolated, oppositely charged
pairs of electrons or muons and requiring the dilepton invariant mass to satisfy
$75<\Mll<105$\GeV.  The \pt\ for each lepton is required to be greater than 20\GeV.
Isolation requirements are relaxed in this channel as the QCD multijet background is
practically eliminated after requiring compatibility with the Z boson mass~\cite{Patrignani:2016xqp}.

\subsubsection{\texorpdfstring{\ptV requirements, \HBB mass reconstruction}{ptV requirements, H to bbbar mass reconstruction}, and event BDT discriminant}

Background events are substantially reduced by requiring significant large
transverse momentum of the reconstructed vector boson, \ptV, or of the Higgs boson candidate~\cite{PhysRevLett.100.242001}.
In this kinematic region, the \Vvar and $\PH$ bosons recoil from each other
with a large azimuthal opening angle,  $\Delta\phi(\mathrm{V,H})$, between them.
Different \ptV regions are selected for each channel.  Because of different signal and background
content, each of these regions has different sensitivity and the analysis is performed separately in each region.  For the \zerol\ channel, a single
region requiring $\MPT>170$\GeV is studied.
The \onel\ channel is also analyzed in a single region, with $\ptV>100$\GeV.
The \twol\  channels consider two regions:  low- and high-\pt regions defined by $50<\ptV<150$\GeV and $\ptV>150$\GeV.

After all event selection criteria described in this section are applied, the
dijet invariant mass resolution of the two \cPqb\ jets from the Higgs boson decay is approximately 15\%, depending on the \pt\ of the reconstructed Higgs boson,
with a few percent shift in the value of the mass peak relative to 125\GeV. The
Higgs boson mass resolution is further improved by applying multivariate
regression techniques similar to those used at the CDF experiment~\cite{1107.3026}
and used for several Run~1 \HBB analyses by ATLAS and CMS~\cite{Chatrchyan:1610290,Aad:2014xzb}.
The regression estimates a correction that is applied after the jet energy
corrections discussed in Section~\ref{sec:hbb_Event_Reconstruction}. It is computed for individual \cPqb\ jets in an attempt to improve
the accuracy of the measured energy with respect to the \cPqb\ quark energy.  To this
end, a BDT is trained on \cPqb\ jets from simulated \ttbar events with inputs
that include detailed jet structure information, which differs in jets from \cPqb\ quarks
from that of jets from light-flavor quarks or gluons.  These inputs include variables
related to several properties of the secondary vertex (when reconstructed), information
about tracks, jet constituents, and other variables related to the energy reconstruction
of the jet.  Because of semileptonic \cPqb\ hadron decays, jets from \cPqb\ quarks
contain, on average, more leptons and a larger fraction of missing energy than jets from
light quarks or gluons.  Therefore, in the cases where a low-\pt lepton is found in the
jet or in its vicinity, the following variables are also included in the regression BDT:
the \pt of the lepton, the $\Delta R$ distance between the lepton and the jet directions,
and the momentum of the lepton transverse to the jet direction.

\begin{figure}
 \centering
    \includegraphics[width=0.48\textwidth]{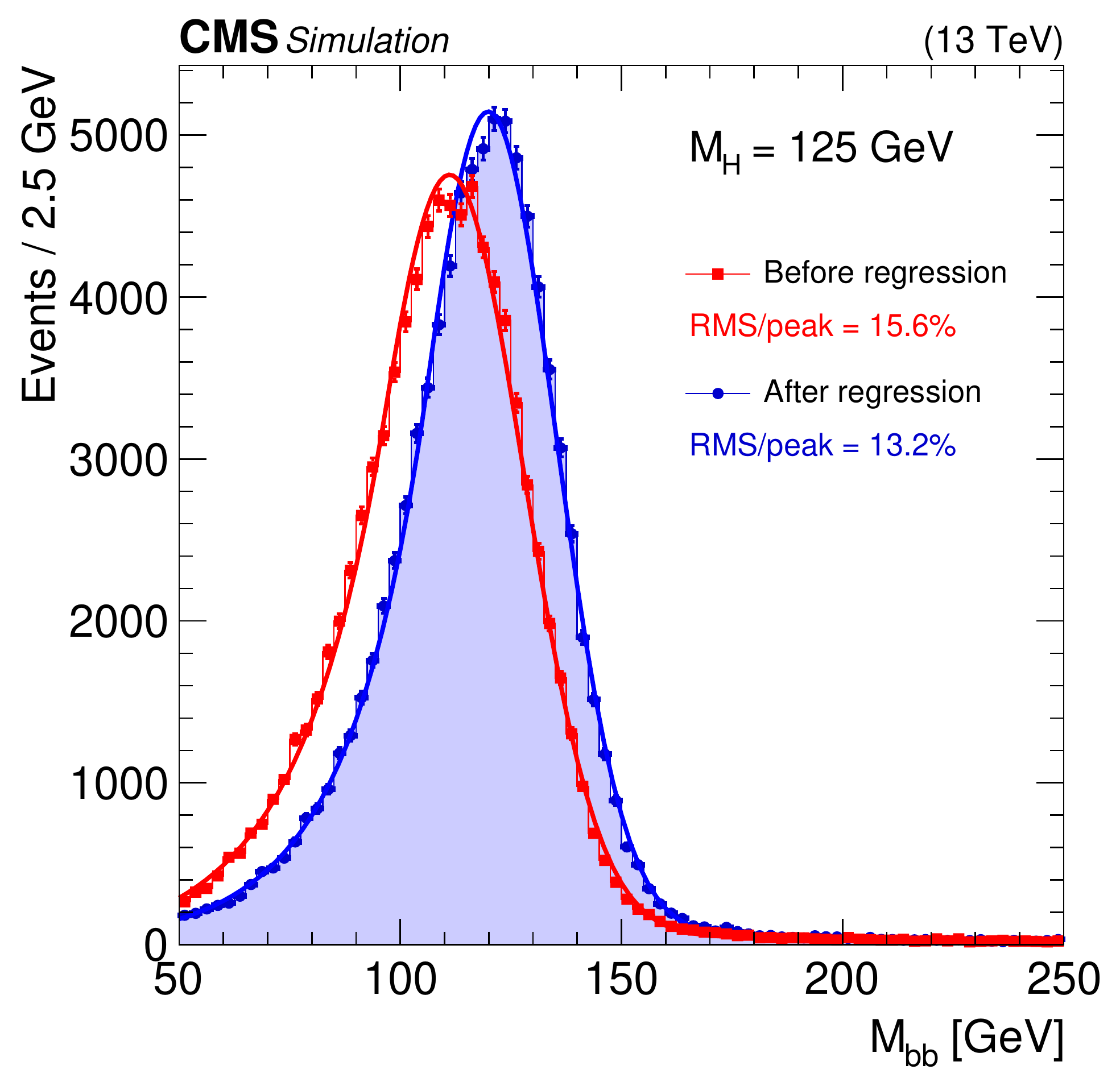}
    \caption{
  Dijet invariant mass distributions for simulated samples of \ZllHbb\ events
    ($m_{\PH} = 125\GeV$), before (red) and after (blue) the energy correction from the
    regression procedure is applied. A sum of a Bernstein polynomial and a
    Crystal Ball function is used to fit the distribution.  The displayed resolutions are
    derived from the peak and RMS of the Gaussian core of the Crystal Ball function.
    }
    \label{fig:regression_VV_VH}

\end{figure}

For the three channels under consideration, the \HBB mass resolution, measured on simulated signal samples  when the regression-corrected jet energies are used, is in the 10--13\% range, and it depends on the \pt\ of the reconstructed Higgs boson.
Averaging over all channels, the improvement in mass resolution is approximately 15\%, resulting in an increase of about 10\% in the sensitivity of the analysis. The performance of these corrections
is shown in Fig.~\ref{fig:regression_VV_VH} for simulated samples of
\ZllHbb\ events. The validation of the technique in data is done using
the \ptll/\ptjj\ distribution in
samples of $\cPZ\to\ell\ell$ events containing two \cPqb-tagged jets,
and using the reconstructed top quark mass distribution in the lepton+jets final state
in $\ttbar$-enriched samples.
After the jets are corrected, the root-mean-square value of both distributions decreases, the peak value of the \ptll/\ptjj\ distribution is shifted
closer to 1.0, and the peak value of the reconstructed top quark mass gets closer to the top quark mass.
These distributions show good agreement between data and the simulated samples before and
after the regression correction is applied.  Importantly, the reconstructed dijet invariant
mass distributions for background processes do not develop a peak structure when the
regression correction is applied to the selected b-tagged jets in the event.

As mentioned above, to help separate signal from background in the signal region, an event BDT discriminant is trained using
simulated samples for signal and all background processes. The set of event input variables used,
listed in Table~\ref{tab:BDTvars}, is chosen by iterative optimization from a larger
number of potentially discriminating variables. Among the most discriminating variables
for all channels are the dijet invariant mass distribution, \Mjj, the number of
additional jets, \Naj, the value of CMVA for the jet with the
second-highest CMVA value, CMVA$_{\text{min}}$, and the distance, \dRJJ, between the two jets.

\begin{table*}
\topcaption{Variables used in the training of the event BDT discriminant for the different channels. Jets are counted as additional jets to
those selected to reconstruct the \HBB decay if they
satisfy the following:
$\pt>30\GeV$ and $\abs{\eta} < 2.4$ for the \zero\
and \twol\ channels,
and $\pt>25\GeV$ and $\abs{\eta} < 2.9$ for the \onel\ channel.
}
\label{tab:BDTvars}
\centering
\cmsTabResize
{
\begin{tabular}{llccc}
\hline
Variable & Description & \zerol & \onel & \twol \\
\hline
\Mjj & dijet invariant mass                                                & \xmark  & \xmark & \xmark \\
\ptjj & dijet transverse momentum                                          & \xmark  & \xmark & \xmark \\
$\pt(\mathrm{j_1})$, $\pt(\mathrm{j_2})$ & transverse momentum of each jet & \xmark  &       & \xmark \\
\dRJJ & distance in $\eta$--$\phi$ between jets                            &        &       & \xmark \\
\dEtaJJ & difference in $\eta$ between jets                                & \xmark  &       & \xmark \\
\dPhiJJ & azimuthal angle between jets                                     & \xmark  &       &       \\
\ptV & vector boson transverse momentum                                    &        & \xmark & \xmark \\
\dphiVH & azimuthal angle between vector boson and dijet directions        & \xmark  & \xmark & \xmark \\
$\ptjj/\ptV$ & \pt\ ratio between dijet and vector boson                   &        &       & \xmark \\
\Mll & reconstructed Z boson mass                                          &        &       & \xmark \\
CMVA$_{\mathrm{max}}$ & value of CMVA discriminant for the jet             & \xmark  &       & \xmark \\
                       & with highest CMVA value                           &        &       &       \\
CMVA$_{\mathrm{min}}$ & value of CMVA discriminant for the jet             & \xmark  & \xmark & \xmark \\
                       & with second highest CMVA value                    &        &       &       \\
CMVA$_{\mathrm{add}}$ & value of CMVA for the additional jet               & \xmark  &       &       \\
                       & with highest CMVA value                           &        &       &       \\
\MPT & missing transverse momentum                                         & \xmark  & \xmark & \xmark \\
$\Delta\phi(\MPTvec\mathrm{, }\mathrm{j})$ & azimuthal angle between \MPTvec\  and closest jet ($\pt>30\GeV$)  & \xmark  &       &       \\
$\Delta\phi(\MPTvec\mathrm{, }\ell)$ & azimuthal angle between \MPTvec\ and lepton                             &        & \xmark &       \\
$m_{\mathrm{T}}$ & mass of lepton \ptvec + \MPTvec                         &        & \xmark &       \\
\mtop & reconstructed top quark mass                                       &        & \xmark &       \\
\Naj & number of additional jets                                           &        & \xmark & \xmark \\
$\pt(\text{add})$ & transverse momentum of leading additional jet          & \xmark  &       &       \\
SA5  & number of soft-track jets with $\pt>5$\GeV                          & \xmark  & \xmark & \xmark \\
\hline
\end{tabular}
}
\end{table*}

\subsection{Background control regions}
\label{sec:hbb_CR}

To help determine the normalization of the main background processes, and
to validate how well the simulated samples model the distributions of
variables most relevant to the analysis, several control regions are
selected in data. Tables~\ref{tab:ZnnControl}--\ref{tab:ZllControl} list the
selection criteria used to define these regions for the \zero, \one, and
\twol\ channels, respectively.
Separate control regions are specified for \ttbar\ production and for the production of
\PW\ and \cPZ\ bosons in association with either predominantly heavy-flavor (HF) or
light-flavor (LF) jets. While some control regions are very pure in their
targeted background process, others contain more than one process.

\begin{table*}
\topcaption{
Definition of the control regions for the \zerol\ channel.
LF and HF refer to light- and heavy-flavor jets.
The values listed for kinematic variables are in units of \GeV, and
for angles in units of radians.  Entries marked with ``\NA'' indicate
that the variable is not used in that region.
}
\label{tab:ZnnControl}
\centering
\begin{tabular}{l c c c c c c}
\hline
Variable                               & \ttbar     & Z+LF       & Z+HF             \\
\hline
V decay category                       &\Wln        &\Znn        &\Znn              \\
$\ptone$                               &$>$60       &$>$60       &$>$60             \\
$\pttwo$                               &$>$35       &$>$35       &$>$35             \\
$\ptjj$                                &$>$120      &$>$120      &$>$120            \\
$\MPT$                                 &$>$170      &$>$170      &$>$170            \\
$\dphiVH$                              &$>$2        &$>$2        &$>$2              \\
\Nal                                   &$\geq1$     &$=$0        &$=$0              \\
\Naj                                   &$\geq2$     &$\leq1$     &$<$1           \\
\Mjj                                   &\NA         &\NA         &$\notin[60-160]$  \\
CMVA$_{\mathrm{max}}$                  &$>$\CMVAM   &$<$\CMVAM   &$>$\CMVAT         \\
CMVA$_{\mathrm{min}}$                  &$>$\CMVAL  &$>$\CMVAL    &$>$\CMVAL        \\
$\Delta\phi(\mathrm{j},\MPTvec)$          &\NA         &$>$0.5      &$>$0.5            \\
$\dphiMtrk$                            &\NA         &$<$0.5      &$<$0.5            \\
$\min\Delta\phi(\mathrm{j},\MPTvec)$      &$<\pi/2$    &\NA         &\NA               \\
\hline
\end{tabular}
\end{table*}

\begin{table*}
\topcaption{Definition of the control regions for the \onel\ channels.
  The HF control region is divided into low- and high-mass ranges
  as shown in the table.  The significance of \MPT, \METsig, is \MPT\ divided by
  the square root of the scalar sum of jet \pt where jet $\pt>30$\GeV. The
  values listed for kinematic variables are in units of \GeV, except for \METsig whose units are
  $\sqrt{\GeV}$.  For angles units are radians.  Entries marked
  with ``\NA'' indicate that the variable is not used in that region.
}
\label{tab:WlnControl}
\centering
{
\begin{tabular}{lccc} \hline
    Variable                & \ttbar        & W+LF                & W+HF            \\
    \hline
    $\ptone$                & $>$25         & $>$25               & $>$25           \\
    $\pttwo$                & $>$25         & $>$25               & $>$25           \\
    \ptjj                   & $>$100        & $>$100              & $>$100          \\
    \ptV                    & $>$100        & $>$100              & $>$100          \\
    CMVA$_{\text{max}}$   & $>$\CMVAT     & [\CMVAL,\CMVAM]     & $>$\CMVAT        \\
    \Naj                    & $>$1          & \NA                  & $=$0            \\
    \Nal                    & $=$0          & $=$0                & $=$0            \\
    \METsig                 & \NA            & $>$2.0              & $>$2.0          \\
    $\Delta\phi(\MPTvec,\ell)$ & $<$2          & $<$2                & $<$2            \\
    $\Mjj$                  & $<$250        & $<$250           & $<$90, $[150,250]$  \\
    \hline
\end{tabular}
}
\end{table*}

\begin{table*}[tbp]
\topcaption{Definition of the control regions for the \twol\ channels. The
same selection is used for both the low- and high-\ptV regions.
The values listed for kinematic variables are in units of \GeV and for angles
in units of radians.  Entries marked
with ``\NA'' indicate that the variable is not used in that region.}
\label{tab:ZllControl}
\centering
\begin{tabular}{lccc} \hline
 Variable                & \ttbar                             & Z+LF              & Z+HF             \\ \hline
\ptV                     & $[50,150]$,$>$150               & $[50,150]$,$>$150 & $[50,150]$,$>$150\\
CMVA$_{\text{max}}$    & $>$\CMVAT                         & $<$\CMVAL         &   $>$\CMVAT      \\
CMVA$_{\text{min}}$    & $>$\CMVAL                         & $<$\CMVAL        &   $>$\CMVAL          \\
\MPT                     & \NA                                  & \NA             &   $<$60           \\
\dphiVH                  & \NA                                  & $>$2.5          &   $>$2.5         \\
\Mll                     &  $\notin[0,10]$, $\notin[75,120]$  & $[75,105]$        &   $[85,97]$    \\
\Mjj                     & \NA                                & $[90,150]$  &   $\notin[90,150]$ \\
\hline
\end{tabular}
\end{table*}

Different background processes feature specific b jet compositions, e.g. two genuine
b jets for \ttbar\ and {\Vvar}+bb, one genuine b jet for {\Vvar}+b, no genuine b jet for {\Vvar}+udscg.
This characteristic, together with their different kinematic distributions, results in distinct \CMVAmin\ distributions that serve to extract the normalization scale factors of the various simulated background samples when fit to data in conjunction with the BDT distributions in the signal region to search for a possible \VH\ signal.
 In this signal-extraction fit, discussed further in Section~\ref{sec:hbb_Results}, the shape and normalization of these distributions are allowed to vary,
 for each background component, within the systematic and statistical uncertainties described in Section~\ref{sec:hbb_Uncertainties}. These uncertainties are treated as independent nuisance parameters. The simulated samples for the {\Vvar}+jets processes are split into independent subprocesses according to the number of MC generator-level jets (with $\pt >20$\GeV and $\abs{\eta}<2.4$)
containing at least one \cPqb\ hadron. Table~\ref{tab:SFs2016}
lists the scale factors obtained from the fit.
These account not only for possible cross section discrepancies, but also for potential residual differences in the selection efficiency of the different objects in the detector. Scale factors obtained from a similar fit to the control regions alone are consistent with those in Table~\ref{tab:SFs2016}.
Given the significantly different event selection criteria, each channel probes different kinematic and topological features of the same background processes and variations in the value of the scale factors across channels are to be expected.

\begin{figure*}[tbhp]
 \centering
  \includegraphics[width=\cmsFigWidthVI]{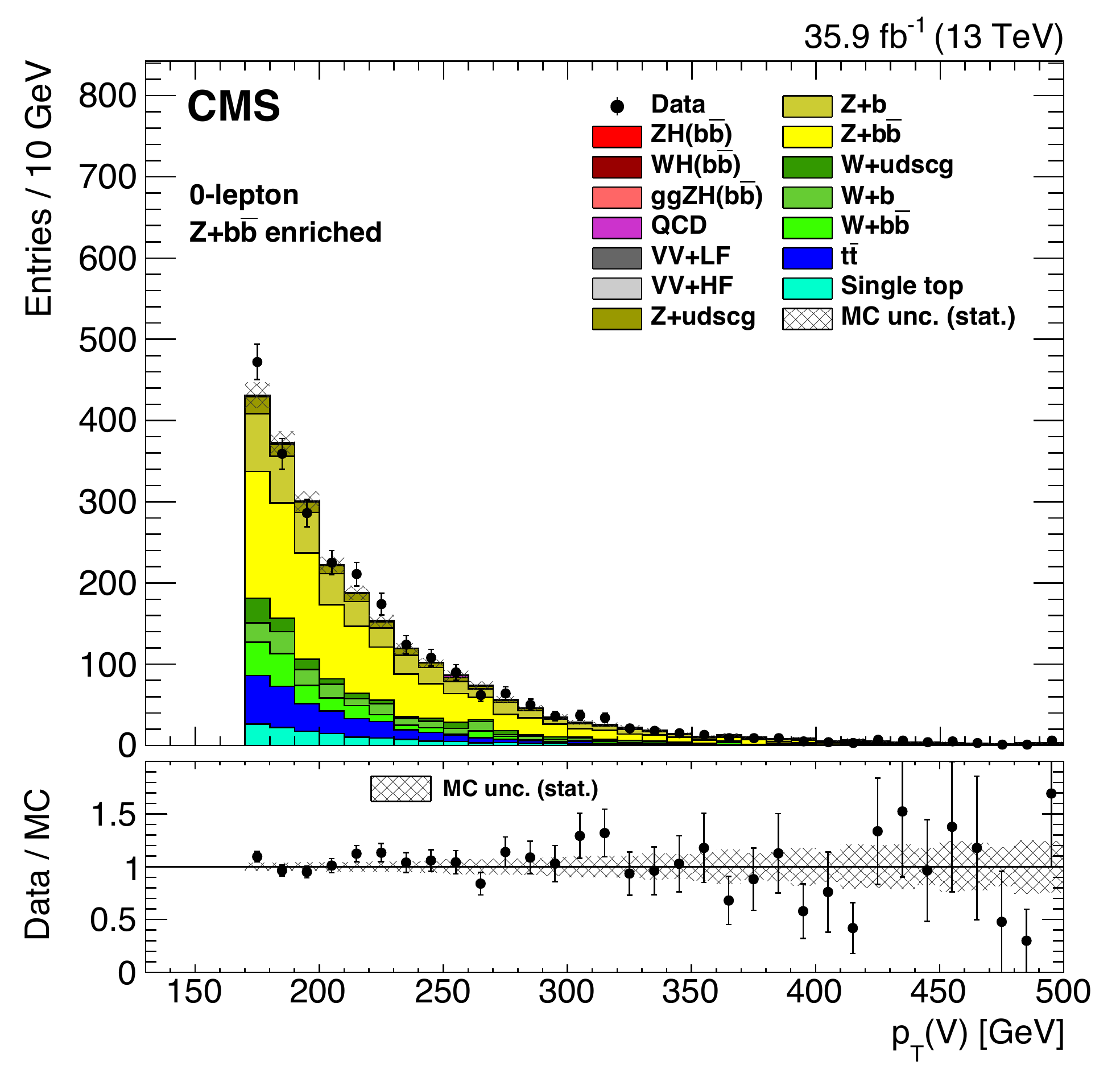}
  \includegraphics[width=\cmsFigWidthVI]{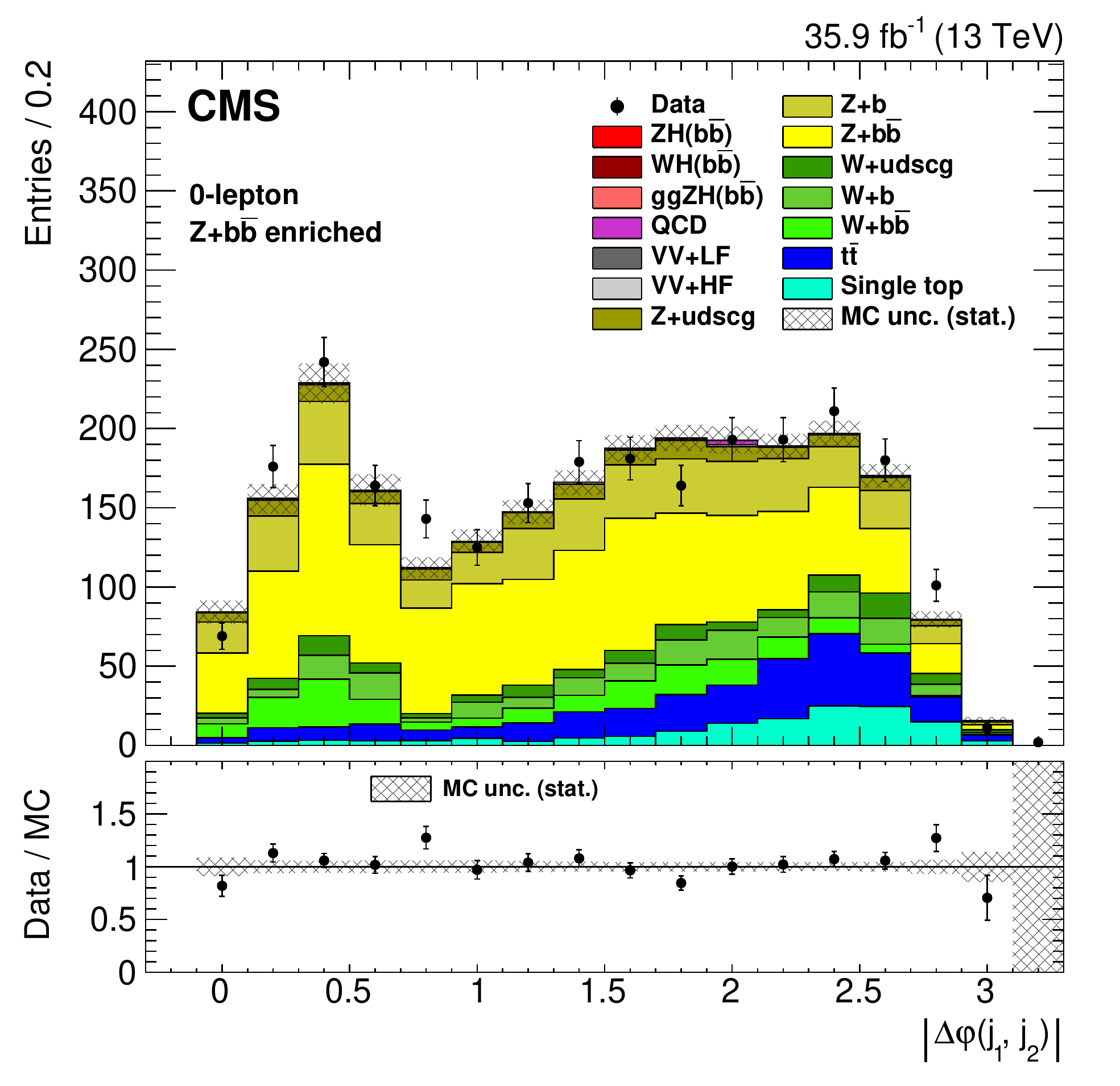}
  \includegraphics[width=\cmsFigWidthVI]{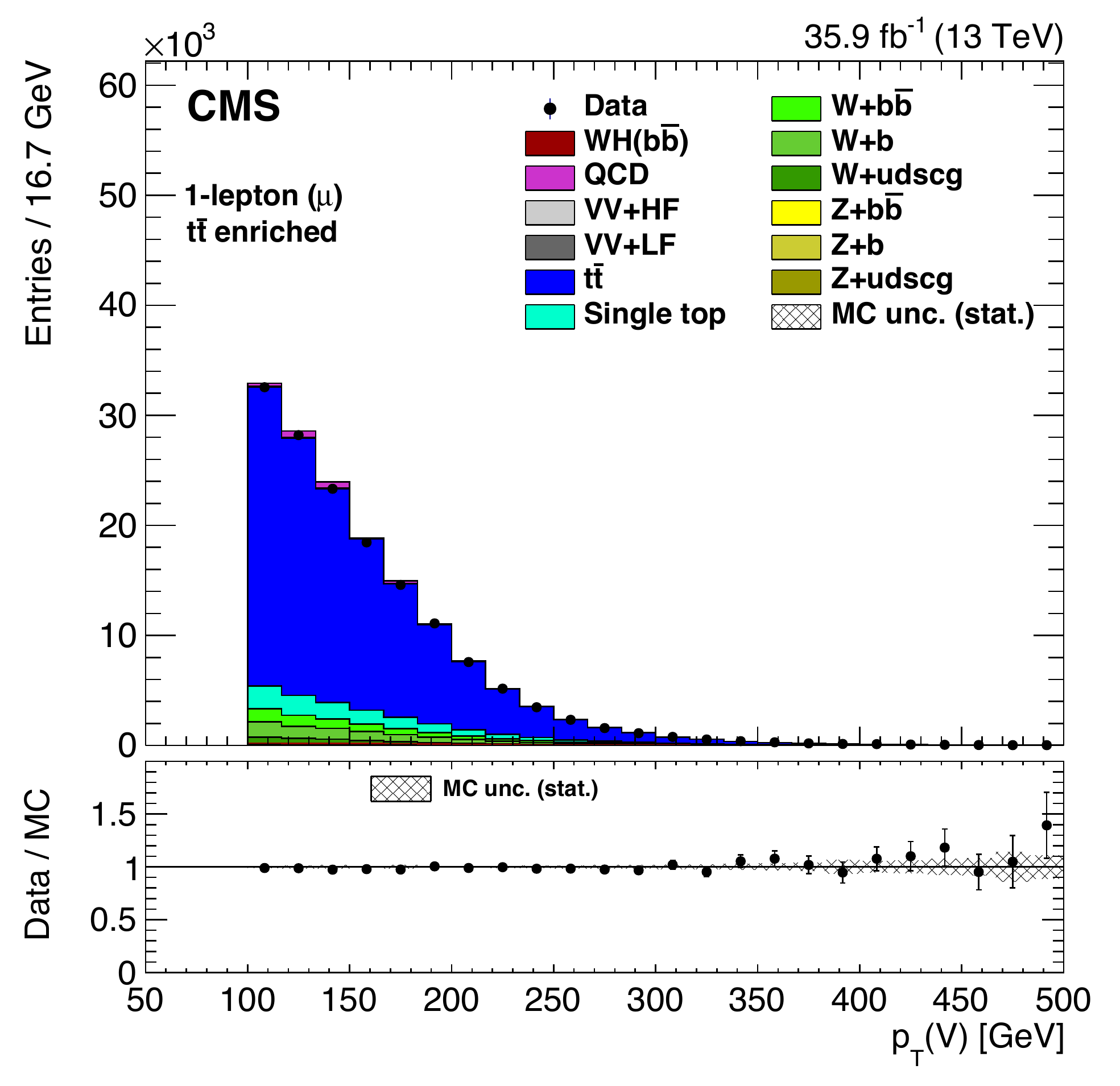}
  \includegraphics[width=\cmsFigWidthVI]{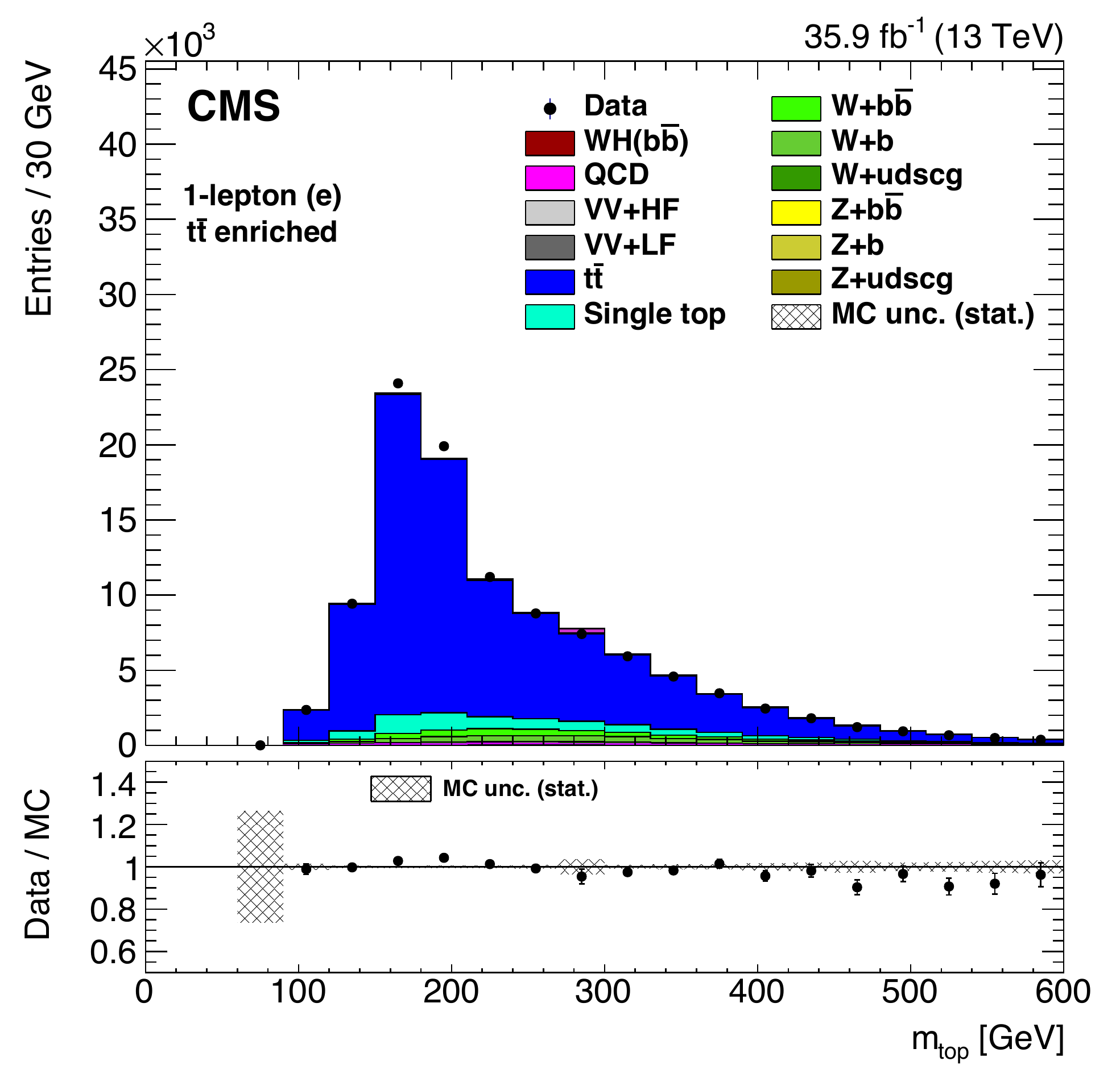}
  \includegraphics[width=\cmsFigWidthVI]{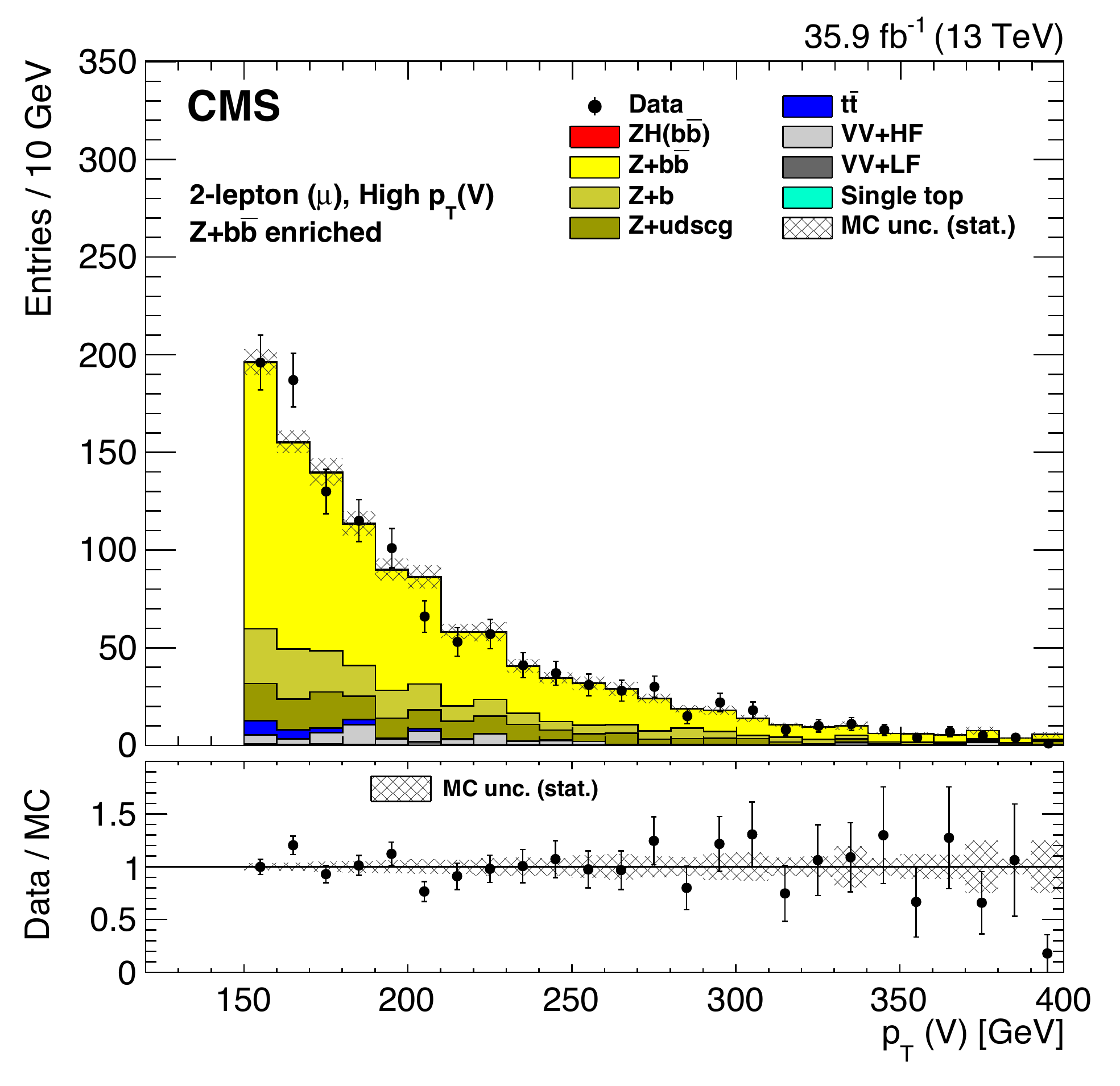}
  \includegraphics[width=\cmsFigWidthVI]{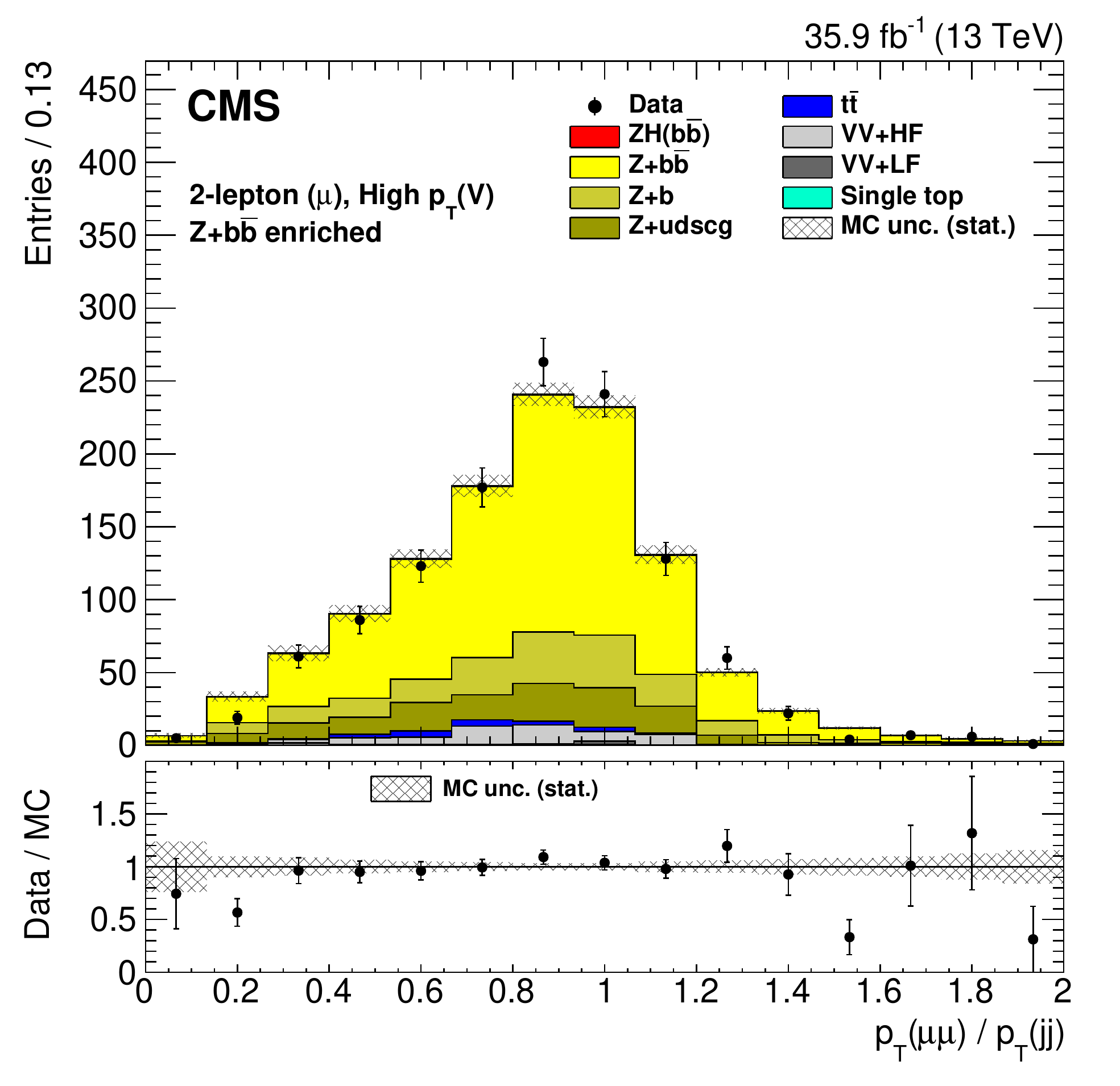}
    \caption{
      Examples of distributions for variables in the simulated
      samples and in data for different control regions and for different channels after applying the
      data/MC scale factors in Table~\ref{tab:SFs2016}.  The top row of plots is from the \zerol\
      Z+HF control region.  The middle row shows variables in the \onel\ \ttbar\ control region.
      The bottom row shows variables in the \twol\ Z+HF control region.  The plots on the left are
      always \ptV.  Plots on the right show a key variable that is validated in that control region.  These variables are,
      from top to bottom, the azimuthal angle between the two jets that comprise the Higgs boson, the
      reconstructed top quark mass, and the ratio of \ptV and \ptjj.
    }
    \label{fig:control_regions_ex}
\end{figure*}

\begin{figure*}[tbhp]
  \centering
    \includegraphics[width=\cmsFigWidthVI]{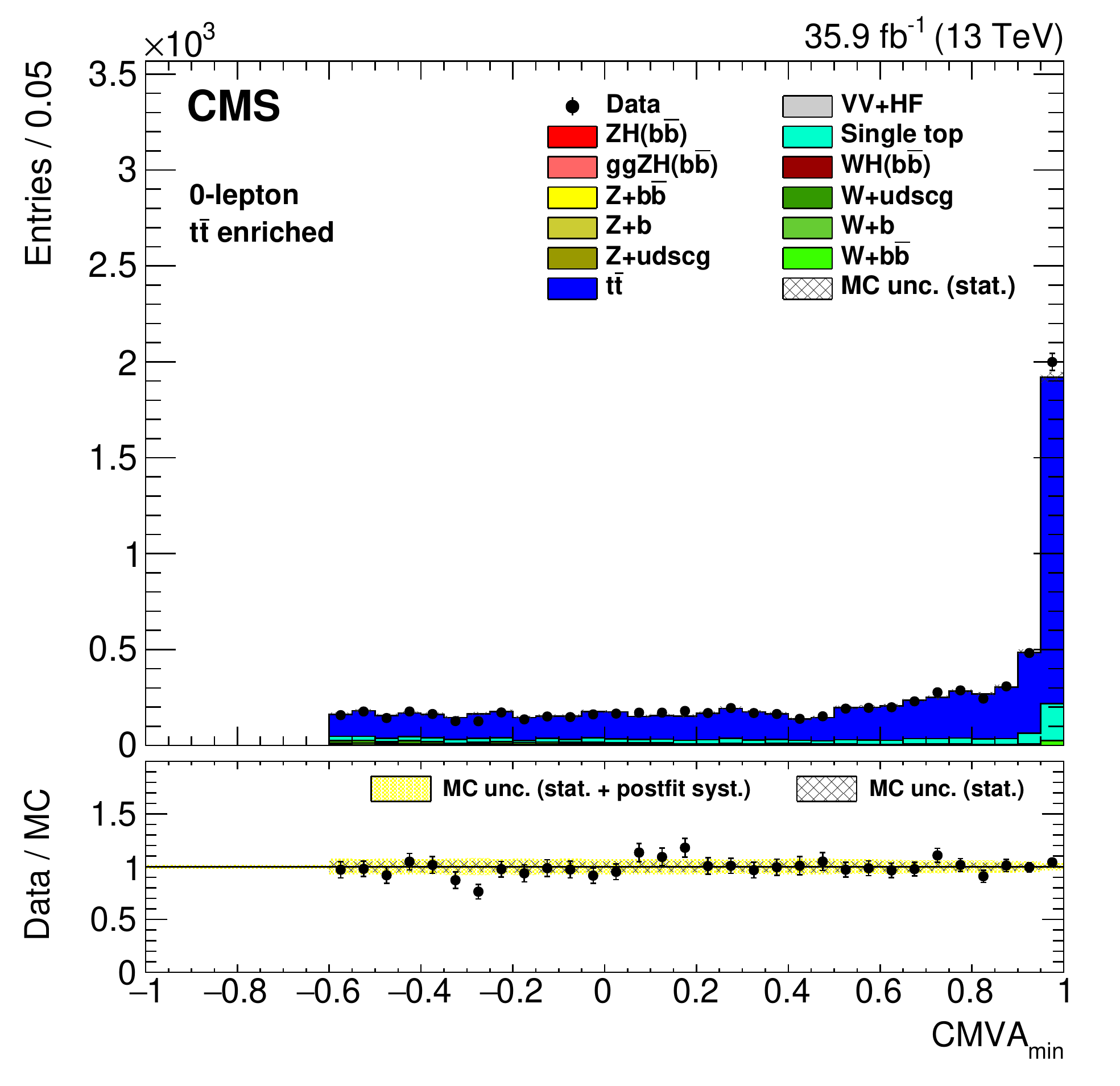}
    \includegraphics[width=\cmsFigWidthVI]{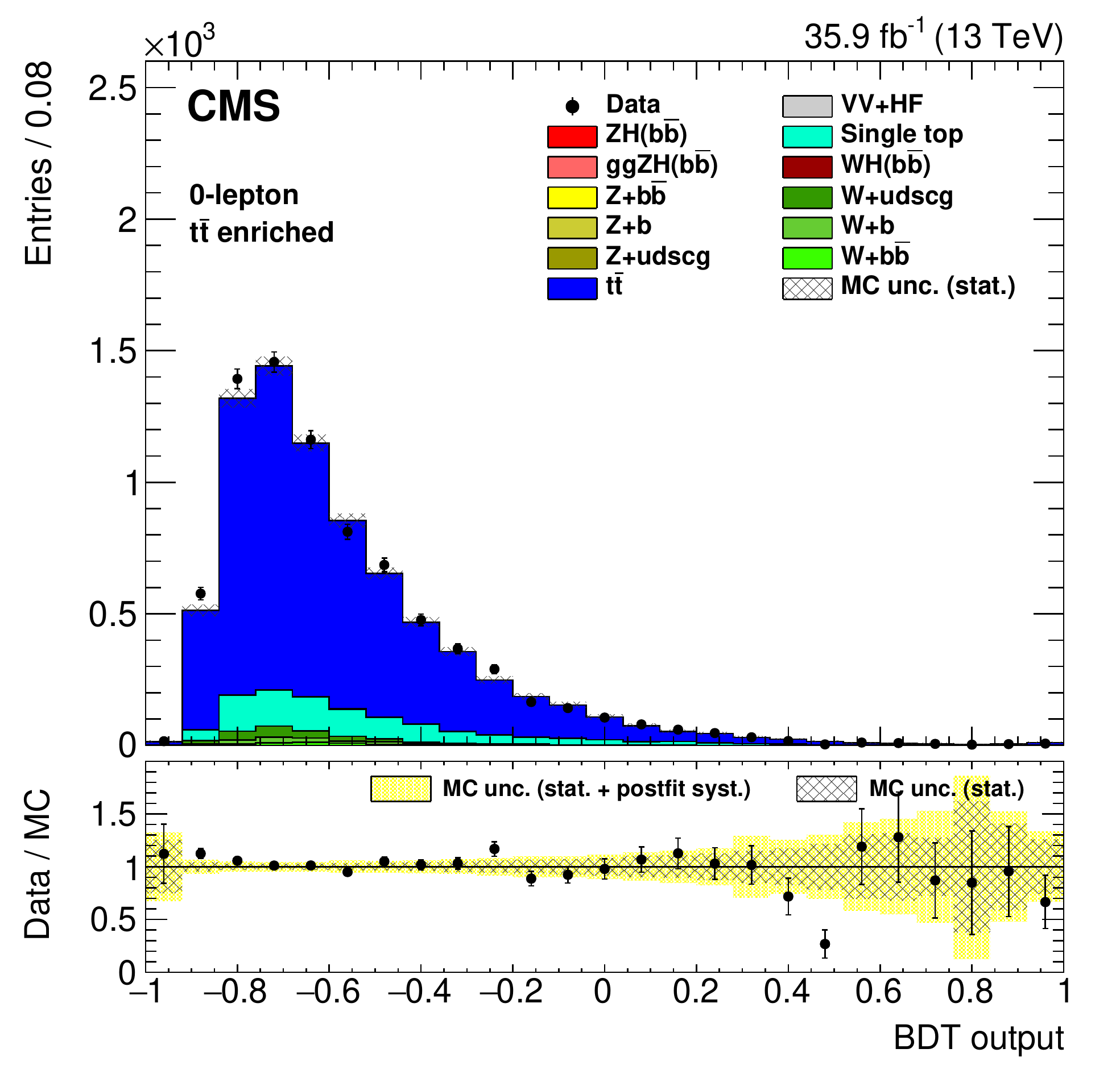}
    \includegraphics[width=\cmsFigWidthVI]{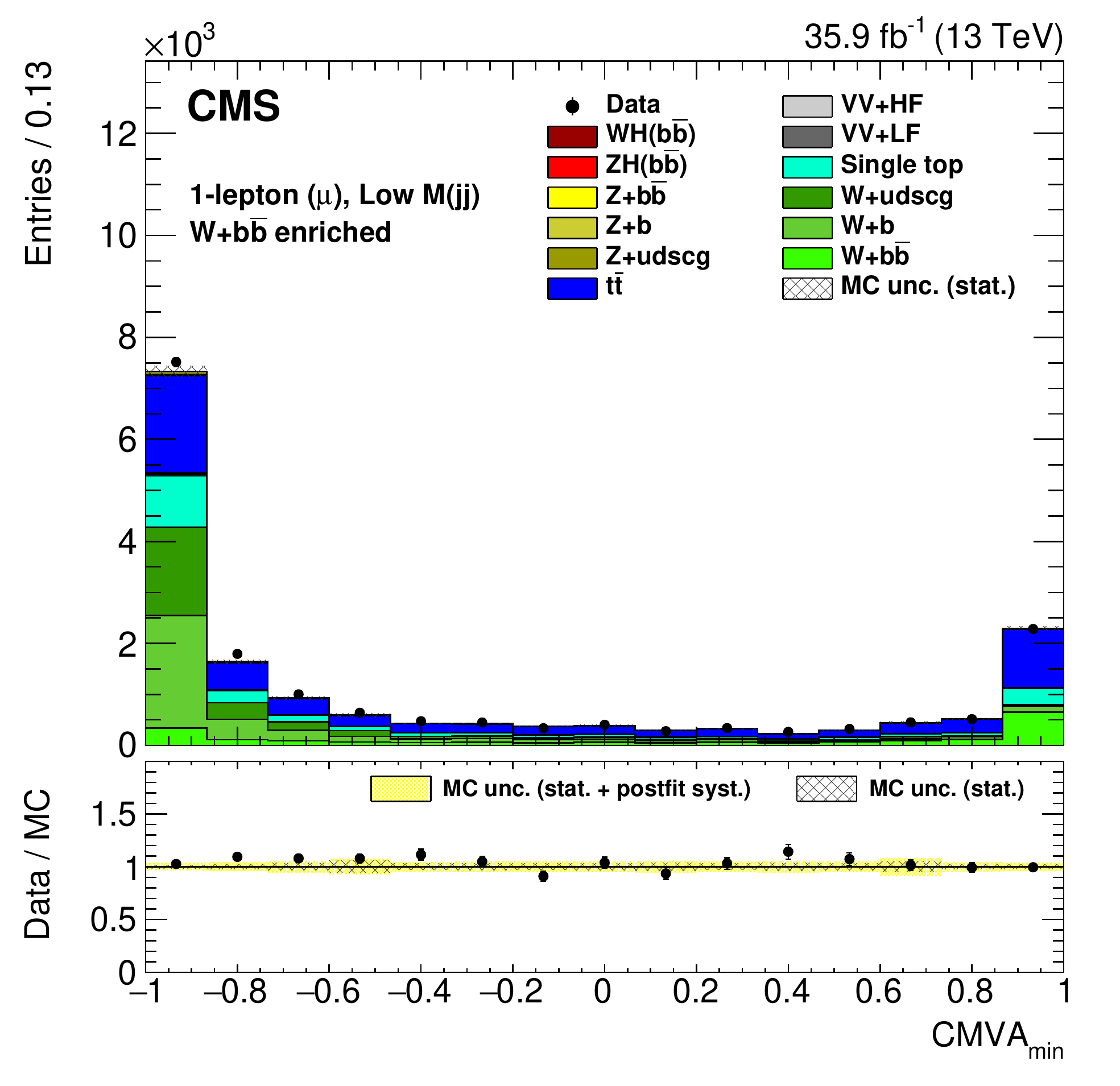}
    \includegraphics[width=\cmsFigWidthVI]{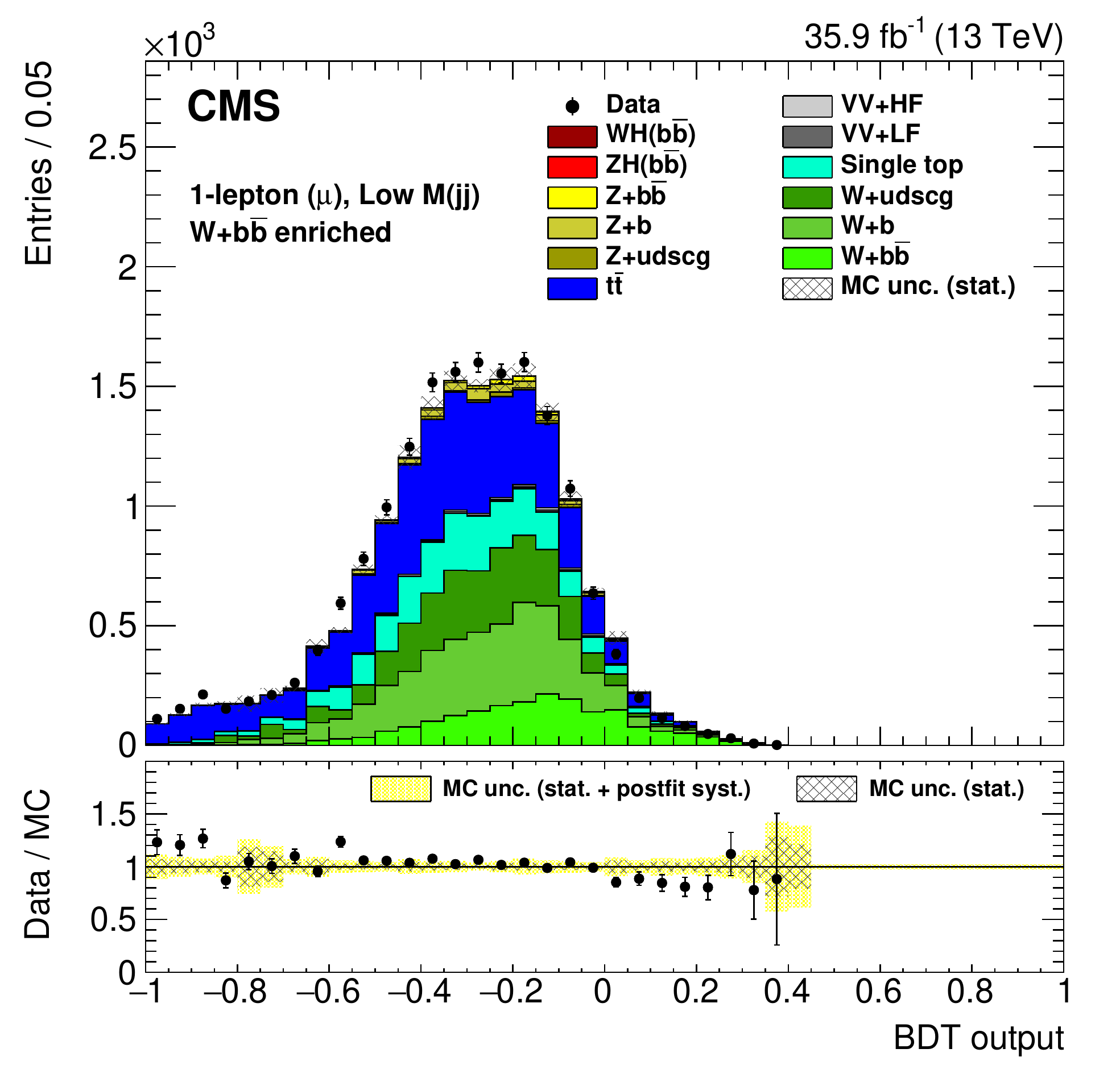}
    \includegraphics[width=\cmsFigWidthVI]{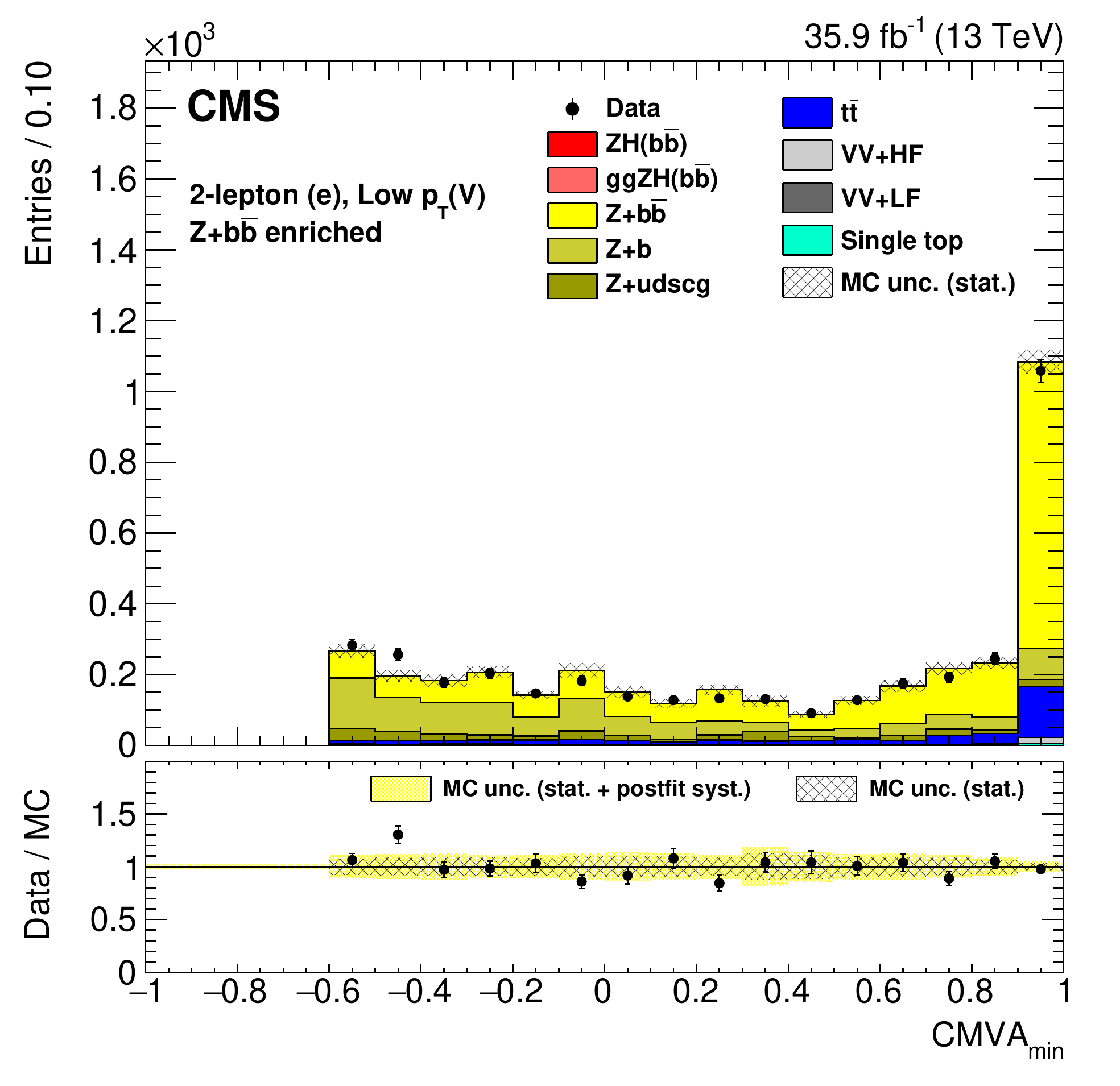}
    \includegraphics[width=\cmsFigWidthVI]{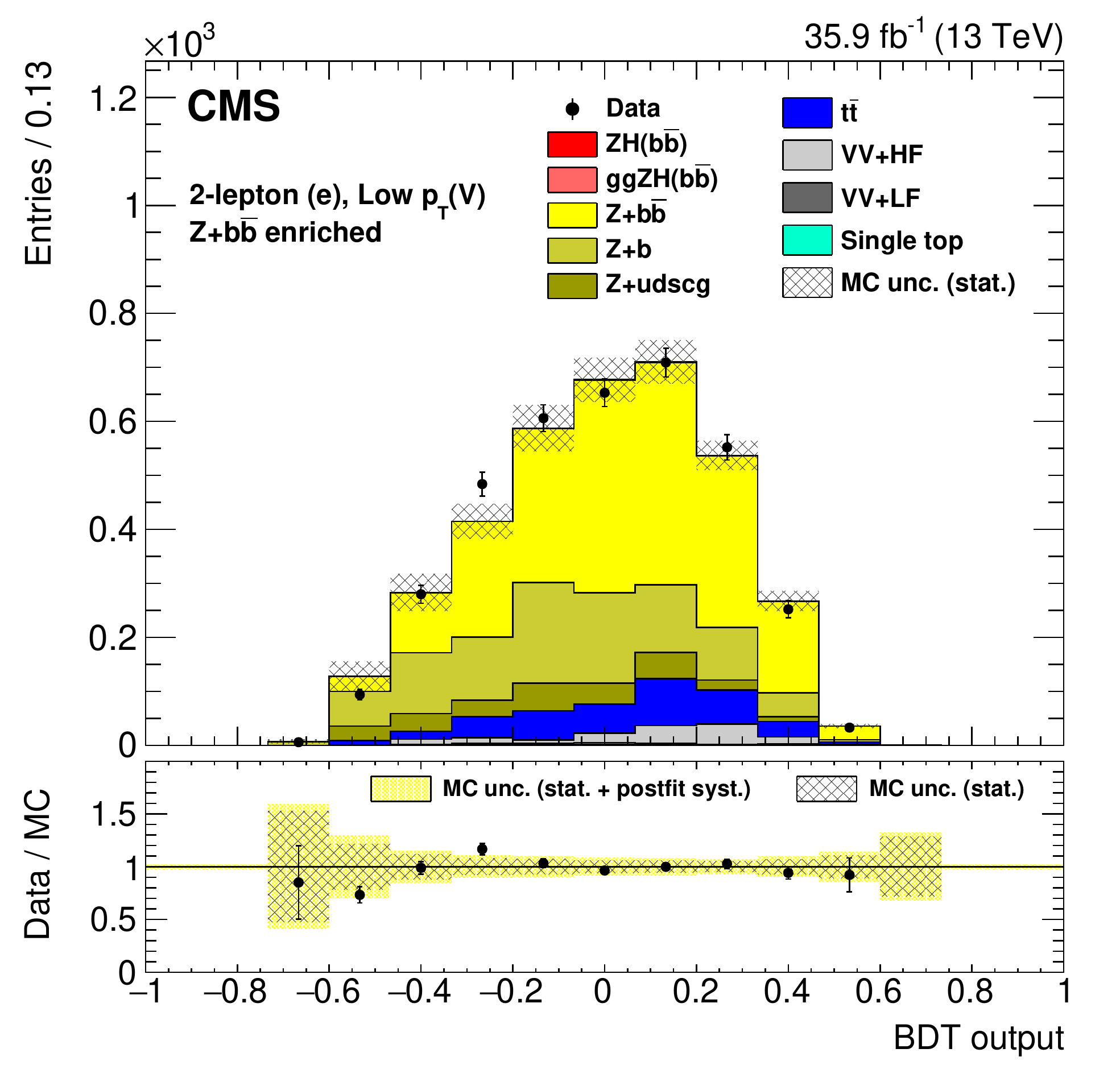}
    \caption{ Distributions in control regions after simulated samples are fit to the data in the
    signal extraction fit.  On the left are examples of \CMVAmin\ distributions, while on the
    right are corresponding event BDT distributions of the same control regions as the plots
    on the left.  Note that these BDT distributions are not part of the fit and are primarily for validation.  The
    control regions shown from top to bottom are: \ttbar for the \zerol\ channel, low-mass HF for the single-muon channel, and HF for the dielectron channel.}
    \label{fig:control_regions_cmvamin_bdt}
\end{figure*}

Figure~\ref{fig:control_regions_ex} shows \ptV distributions together with examples of distributions for
variables in different control regions and for different channels after the scale factors in Table~\ref{tab:SFs2016} have been applied to the corresponding simulated samples.  Figure~\ref{fig:control_regions_cmvamin_bdt} shows examples of \CMVAmin\ and event BDT distributions, also for different control regions and for different channels, where not only the scale factors are applied but also the shapes of the distributions are allowed to vary according to the treatment of systematic uncertainties from all nuisances in the signal-extraction fit. These BDT distributions are from control regions and do not participate in that fit.
The signal region BDT distributions used in the fit are presented in Section~\ref{sec:hbb_Results}.

\begin{table*}[tbp]
\topcaption{Data/MC scale factors for each of the main background processes in each channel, as obtained from the combined signal-extraction fit to control and signal region distributions described in Section~\ref{sec:hbb_Results}.
Electron and muon samples in the \one\ and \twol\ channels are fit simultaneously to determine average scale factors.
The same scale factors for W+jets processes are used  for the \zero\ and \onel\ channels.
}
\label{tab:SFs2016}
\centering
{
\begin{tabular}{lcccccc} \hline
   Process       & \zerol\   & \onel\  & \twol\ low-\ptV  & \twol\ high-\ptV         \\ \hline

W0b      & $1.14 \pm 0.07$  & $1.14 \pm 0.07$  & \NA               & \NA                \\
W1b      & $1.66 \pm 0.12$  & $1.66 \pm 0.12$  & \NA               & \NA                \\
W2b      & $1.49 \pm 0.12$  & $1.49 \pm 0.12$  & \NA               & \NA                \\
Z0b      & $1.03 \pm 0.07$  & \NA               & $1.01 \pm 0.06$  & $1.02 \pm 0.06$   \\
Z1b      & $1.28 \pm 0.17$  & \NA               & $0.98 \pm 0.06$  & $1.02 \pm 0.11$   \\
Z2b      & $1.61 \pm 0.10$  & \NA               & $1.09 \pm 0.07$  & $1.28 \pm 0.09$   \\
\ttbar   & $0.78 \pm 0.05$  & $0.91 \pm 0.03$  & $1.00 \pm 0.03$  & $1.04 \pm 0.05$   \\
    \hline
  \end{tabular}
}
\end{table*}

In inclusive vector boson samples, selected for this analysis, the \ptV spectrum in data is observed to be
softer than in simulated samples, as expected from higher-order electroweak corrections
to the production processes~\cite{NLO_plus_QCD_weight}. The events in all three channels
are re-weighted to account for the electroweak corrections to \ptV.  The correction is negligible
for low \ptV but is sizable at high \ptV, reaching 10\% near 400\GeV.

After these corrections, a residual discrepancy in \ptV between data and simulated samples
is observed in some control regions.  In the \zerol\ channel, \ttbar\
samples are re-weighted as a function of the generated top quark's \pt\ according to the
observed discrepancies in data and simulated samples in differential top quark cross section
measurements~\cite{PhysRevD.95.092001}.  This re-weighting resolves the discrepancy
in \ptV in \ttbar\ control regions.  In the \onel\ channel, additional corrections are
needed for W+jets samples, and corrections are derived from the data in \onel\ control regions for
these processes:  \ttbar,  {\PW}+udscg, and the sum of {\PW}+b, {\PW}+bb, and single top quark backgrounds.
A re-weighting of simulated events in \ptV is derived for each, such that the shape of the sum of simulated
processes matches the data.  The correction functions are extracted through a
simultaneous fit of linear functions in \ptV.  The uncertainties in the fit parameters are
used to assess the systematic uncertainty.  The \ptV spectra resulting from re-weighting
in either the top quark \pt\ or \ptV are equivalent.

The {\Vvar}+jets LO simulated samples are used in the analysis because, due to computing resource limitations, considerably more events are available than for the NLO samples. A normalization K factor is applied to the LO samples to account for the difference in cross sections.
Kinematic distributions between the two samples are found to be consistent after
matching the LO distribution of the pseudorapidity separation $\Delta\eta(\mathrm{jj})$ between the two \HBB jet candidates to the NLO one.
Different corrections are derived depending on whether these two jets are matched to zero, one, or two b quarks. Both the $\Delta\eta(\mathrm{jj})$ distributions of the NLO samples and the corrected LO samples agree well with data in control regions.

\section{Systematic uncertainties}\label{sec:hbb_Uncertainties}

Systematic effects affect the \HBB mass resolution, the shapes of the \CMVAmin\ distributions,
the shapes of the event BDT distributions, and the signal and background yields in the most sensitive region of the BDT distributions. The uncertainties
associated with the normalization scale factors of the simulated samples for
the main background processes have the largest impact on the uncertainty in the fitted signal strength $\mu$. The next largest effects result from the size of the
simulated samples and from uncertainties in correcting mismodeling of kinematic variables,
both in signal and in background simulated samples.  The next group of significant
systematic uncertainties are related to \cPqb\ tagging uncertainties and uncertainties in
jet energy.   All systematic uncertainties considered are listed in Table~\ref{tab:syst} and are described in more detail below, in the same order as they appear in the Table.

\begin{table*}
\topcaption{Effect of each source of systematic uncertainty in the expected signal
strength $\mu$.
The third column shows the uncertainty in $\mu$ from each source
when only that particular source is considered. The last column shows the
percentage decrease in the uncertainty when removing that specific source of
uncertainty while applying all other systematic uncertainties.  Due to
correlations, the total systematic uncertainty is larger
than the sum in quadrature of the individual uncertainties.  The second column
shows whether the source affects only the normalization or both the shape and
normalization of the event BDT output distribution. See text for details.}
\label{tab:syst}
\centering
{
\begin{tabular}{lccc}
\hline
                                              &        &   Individual contribution        & Effect of removal to \\
 Source                                       & Type   &   to the $\mu$ uncertainty (\%)  & the $\mu$ uncertainty (\%)\\  \hline
 Scale factors (\ttbar, V+jets)                & norm.  &  9.4    & 3.5 \\

 Size of simulated samples                    & shape  &  8.1    & 3.1  \\
 Simulated samples' modeling                  & shape  &  4.1    & 2.9 \\
 \cPqb\ tagging efficiency                         & shape  &  7.9    & 1.8  \\
 Jet energy scale                             & shape  &  4.2    & 1.8  \\
 Signal cross sections                        & norm.  &  5.3    & 1.1 \\
 Cross section uncertainties & norm.  &  4.7    & 1.1 \\
 \quad (single-top, VV)\\

 Jet energy resolution                        & shape  &  5.6    & 0.9  \\
 \cPqb\ tagging mistag rate                        & shape  &  4.6    & 0.9  \\
 Integrated luminosity                        & norm.  &  2.2    & 0.9  \\
 Unclustered energy                           & shape  &  1.3    & 0.2  \\
 Lepton efficiency and trigger                & norm.  &  1.9    & 0.1  \\\hline
\end{tabular}
}
\end{table*}

The sizes of simulated samples are sometimes limited. If the statistical uncertainty in the content of certain bins in the BDT distributions for the simulated samples is large, Poissonian nuisance parameters are used in the signal extraction binned-likelihood fit.
These are required mainly in the V+jets samples and are a leading source of systematic uncertainty in the analysis.

The corrections to the \ptV spectra in the \ttbar and W+jets samples are applied per sample
according to the uncertainty in the simultaneous \ptV fit described in
Section~\ref{sec:hbb_CR}.  This
uncertainty on the correction is at most 5\% on the background yield near \ptV of 400\GeV.
The shape difference in the event BDT and \CMVAmin\ distributions between simulations of two event generators are used to account
for imperfect modeling in the nominal simulated samples.  For the {\Vvar}+jets, the difference
between the shapes for events generated with the {\MGvATNLO} MC generator at
LO and NLO is considered as a shape systematic uncertainty.  For the \ttbar\ process, the differences
in the shapes between the nominal sample generated with {\POWHEG} and that obtained from the
{\MCATNLO}~\cite{Frixione:2002ik} generator are considered as shape systematic uncertainties.
Variations on the QCD factorization and renormalization scales and on the PDF choice are considered for the simulated signal and background samples. The scales are varied by factors of 0.5 and 2.0, independently, while the PDF uncertainty effect on the shapes of the BDT distributions is evaluated by using the PDF replicas associated to the NNPDF set~\cite{Ball:2014uwa}.

The \cPqb\ tagging efficiencies and the probability to tag as a b jet a jet originating
from a different flavor (mistag) are measured in heavy-flavor enhanced
samples of jets that contain muons and are applied consistently to jets in signal and
background events. The measured uncertainties for the \cPqb\ tagging scale factors are:
1.5\% per \cPqb-quark tag, 5\% per charm-quark tag, and 10\% per mistagged jet
(originating from gluons and light \cPqu, \cPqd, or \cPqs\ quarks)~\cite{CMS-PAS-BTV-15-001}.
These uncertainties are propagated to the \CMVAmin\ distributions by re-weighting
events.  The shape of the event BDT distribution is also affected by the shape of the CMVA
distributions because \CMVAmin\ is an input to the BDT discriminant.  For the \twol\ channel \CMVAmax\ is
also an input to this discriminant.  The signal strength uncertainty
increases by 8\% and 5\%, respectively, due to \cPqb\ tagging efficiency and mistag scale factor
uncertainties propagated through the CMVA distributions and finally to the event BDT distributions.

The uncertainties in the jet energy scale and resolution have an effect on the shape of the
event BDT output distribution because the dijet invariant mass is a crucial input variable to the BDT discriminant.  The impact of the jet energy scale uncertainty is determined by recomputing the BDT output
distribution after shifting the energy scale up and down by its uncertainty.   Similarly,
the impact of the jet energy resolution is determined by recomputing the BDT output
distribution after increasing or decreasing the jet energy resolution. The uncertainties in jet energy scale and resolution affect not only the jets in the event but also the \MPT, which is recalculated when these variations are applied.
The individual contribution to the increase in signal strength uncertainty is found to be around 6\% for
the jet energy scale and 4\% for the jet energy resolution uncertainty.  The uncertainty
in the jet energy scale and resolution vary as functions of jet \pt\ and $\eta$. For the jet energy scale there
are several sources of uncertainty that are derived and applied independently
as they are fully uncorrelated between themselves~\cite{JETMET:2017},
while for the jet energy resolution a single shape systematic is evaluated.

The total \VH\ signal cross section has been calculated to NNLO+NNLL accuracy in QCD, combined with NLO electroweak
corrections, and the associated systematic uncertainties~\cite{deFlorian:2016spz} include the effect of scale variations and
PDF uncertainties.
The estimated uncertainties in the NLO electroweak
corrections are 7\% for the \WH and 5\% for the \ZH production processes, respectively.
The estimate for the NNLO QCD correction results in an uncertainty of 1\% for the \WH and
4\% for the \ZH production processes, which includes the ggZH contribution.

An uncertainty of 15\% is assigned to the event yields obtained from simulated samples for both single top quark and diboson production.
These uncertainties are
about 25\% larger than those from the CMS measurements
of these processes~\cite{Sirunyan:2016cdg,Khachatryan:2016tgp,Khachatryan:2016txa},
to account for the different kinematic regime in which those measurements are performed.

Another source of uncertainty that affects the \MPT\ reconstruction is the estimate of the energy that is not clustered in jets~\cite{CMS-PAS-JME-16-004}. This affects only the \zero\ and \onel\ channels, with an individual contribution to the signal strength uncertainty of 1.3\%.

Muon and electron trigger, reconstruction, and identification efficiencies in
simulated samples are corrected for differences in data and simulation using samples of
leptonic \cPZ\ boson decays.  These corrections are affected by uncertainties
coming from the efficiency measurement method, the lepton selection, and the
limited size of the Z boson samples. They are measured and propagated as
functions of lepton \pt\ and $\eta$.  The parameters describing the turn-on curve
that parametrizes the \ZnnH\ trigger efficiency as a function of \MPT\
are varied within their statistical uncertainties, and are
also estimated for different assumptions on the methods used to derive the efficiency.  The
total individual impact of these uncertainties on lepton identification and trigger efficiencies on the measured signal strength is about 2\%.

The uncertainty in the CMS integrated luminosity measurement is estimated to be 2.5\%~\cite{CMS-PAS-LUM-17-001}.
Events in simulated samples must be re-weighted such that the distribution of pileup
in the simulated samples matches that estimated in data.  A 5\% uncertainty on pileup re-weighting
is assigned, but the impact of this uncertainty is negligible.

The combined effect of the systematic uncertainties results in a 25\% reduction of
the expected significance for the SM Higgs boson rate.

\section{Results}\label{sec:hbb_Results}

Results are obtained from combined signal and background binned-likelihood fits,
simultaneously for all channels, to both the shape of the output distribution of
the event BDT discriminants in the signal region and to the \CMVAmin\ distributions
for the control regions corresponding to each channel. The BDT discriminants are
trained separately for each channel to search for a Higgs boson with a mass of 125\GeV.
To remove the background-dominated portion of the BDT output distribution, only events with
a BDT output value above thresholds listed in Table~\ref{tab:PreSel} are considered. To
achieve a better sensitivity in the search, this threshold is optimized separately for each channel. In this signal-extraction fit, the shape and normalization of all distributions for signal and for each background
component are allowed to vary within the systematic and statistical uncertainties described in
Section~\ref{sec:hbb_Uncertainties}. These uncertainties are treated as independent nuisance
parameters in the fit. Nuisance parameters, the signal strength, and the scale factors
described in Section~\ref{sec:hbb_CR} are allowed to float freely and are adjusted by the fit.

\begin{figure*}[tbhp]
  \centering
       \includegraphics[width=0.33\textwidth]{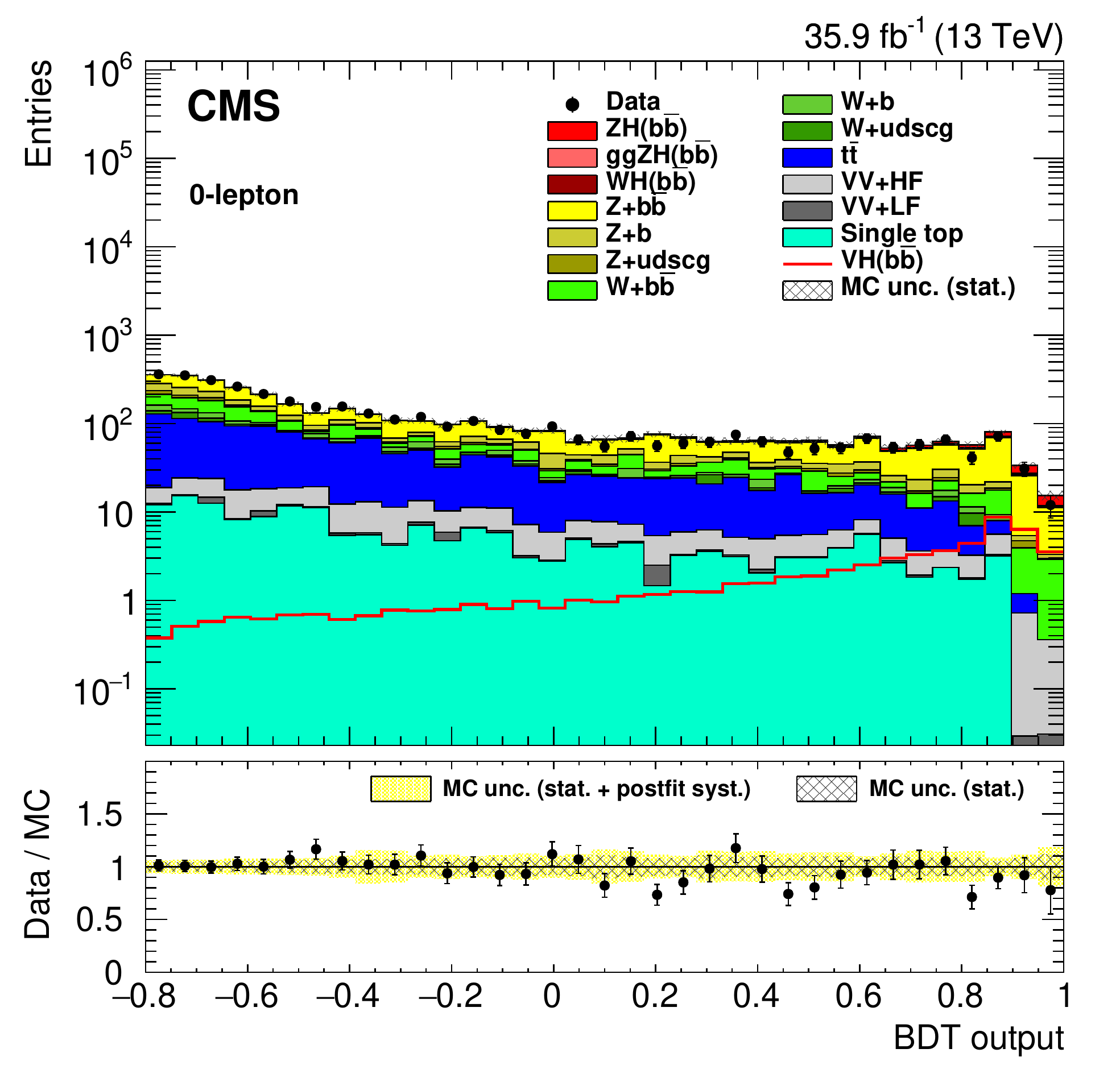} \\
     \includegraphics[width=0.33\textwidth]{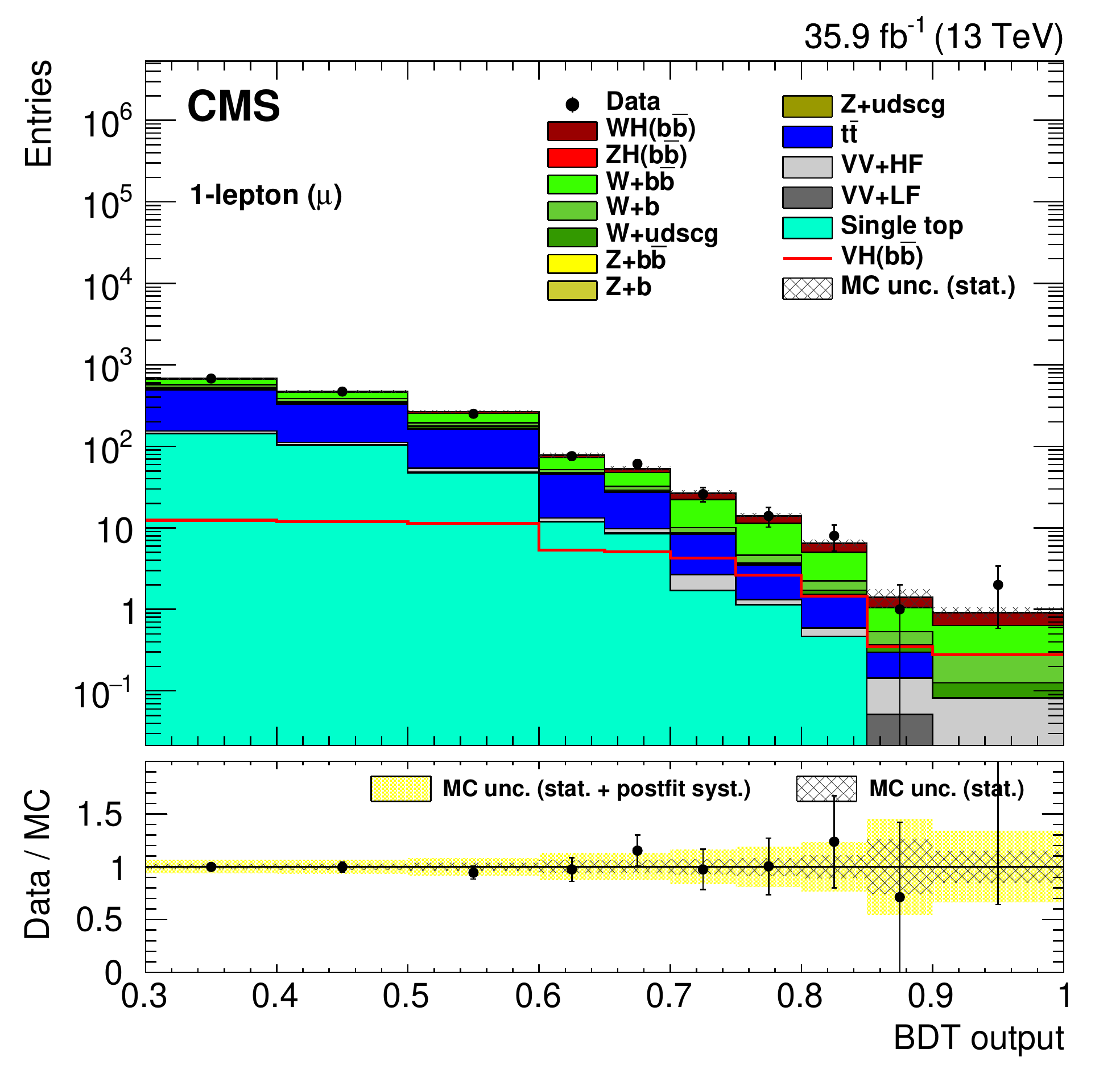}
    \includegraphics[width=0.33\textwidth]{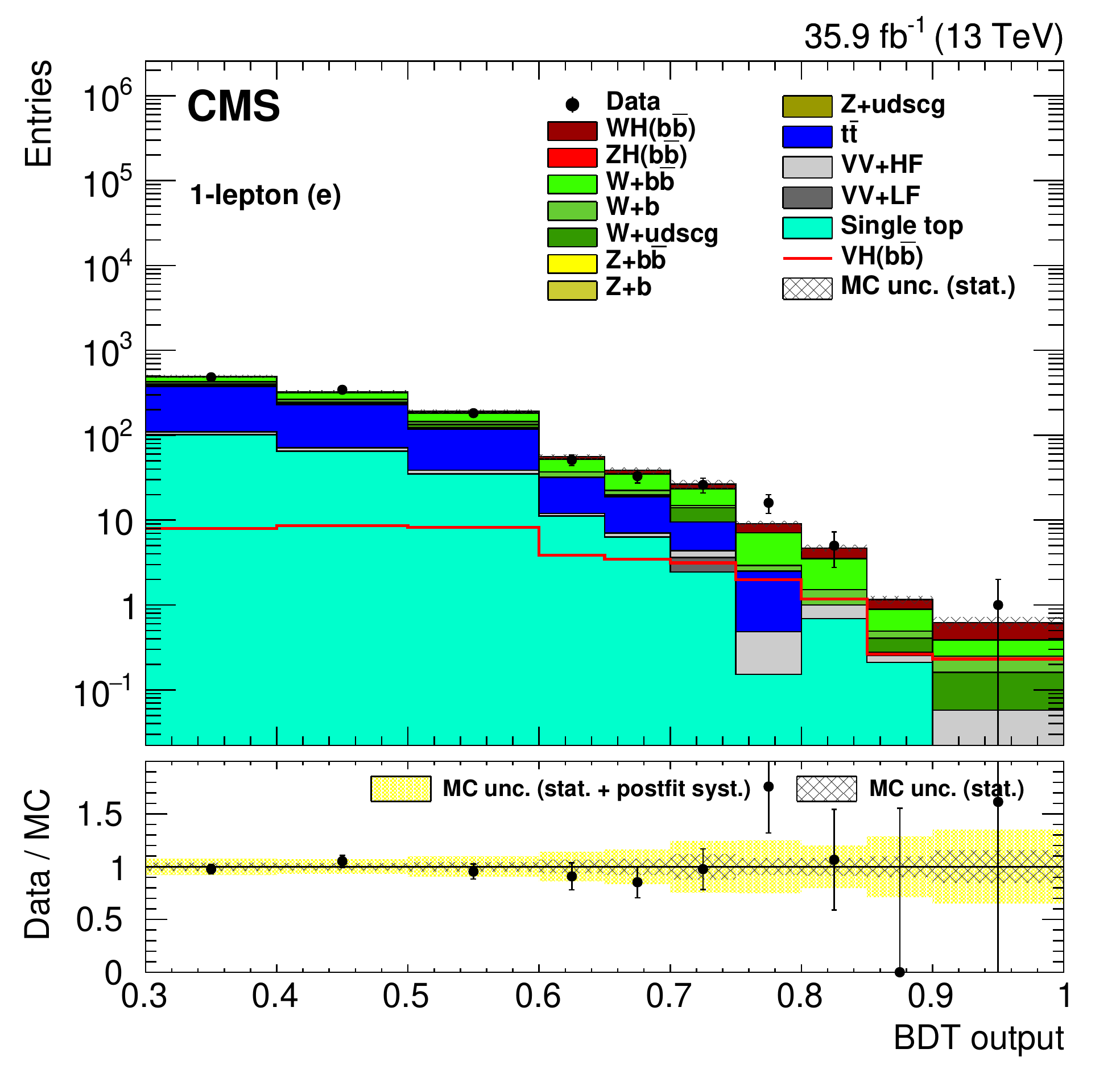}
    \includegraphics[width=0.33\textwidth]{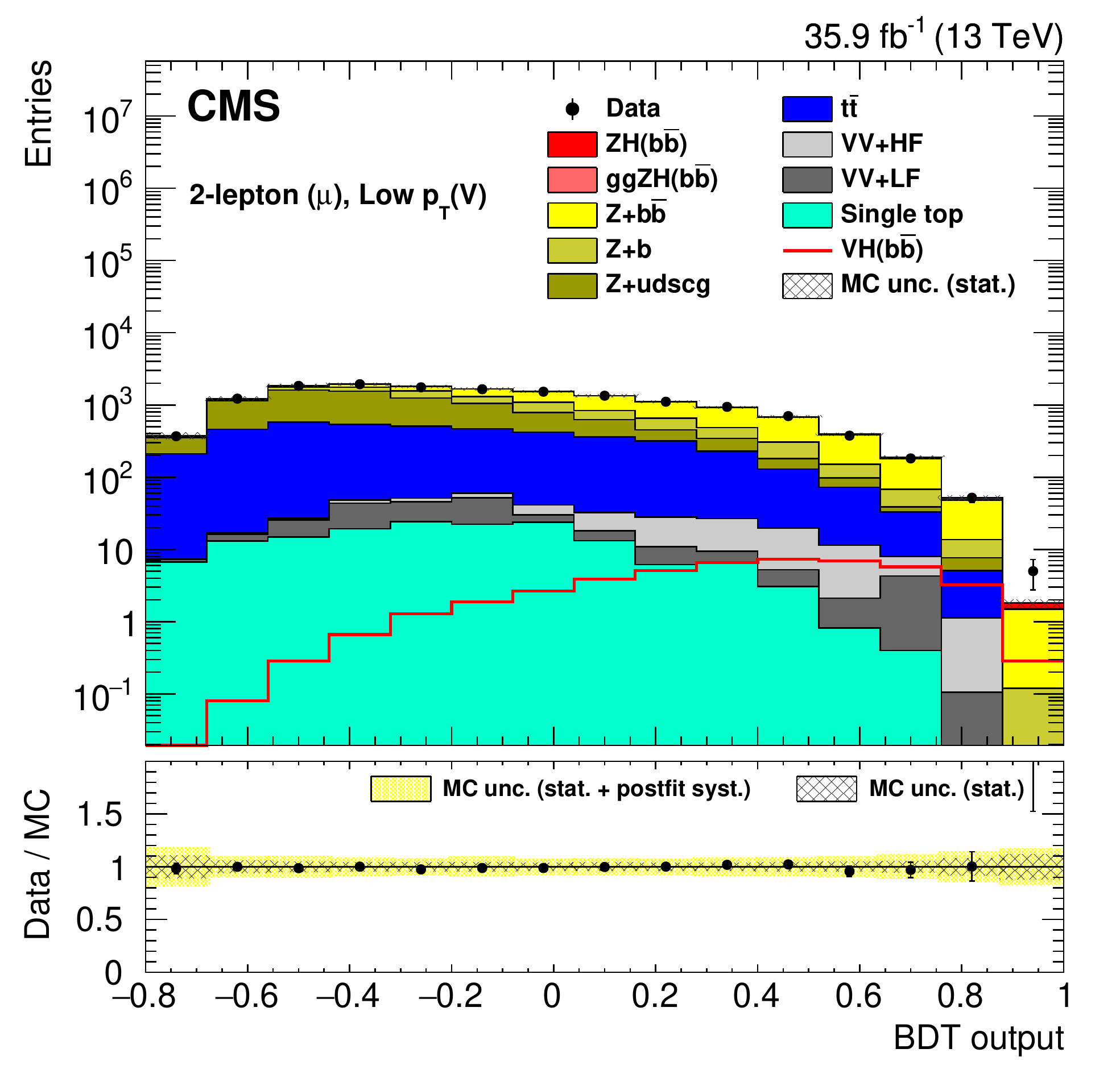}
    \includegraphics[width=0.33\textwidth]{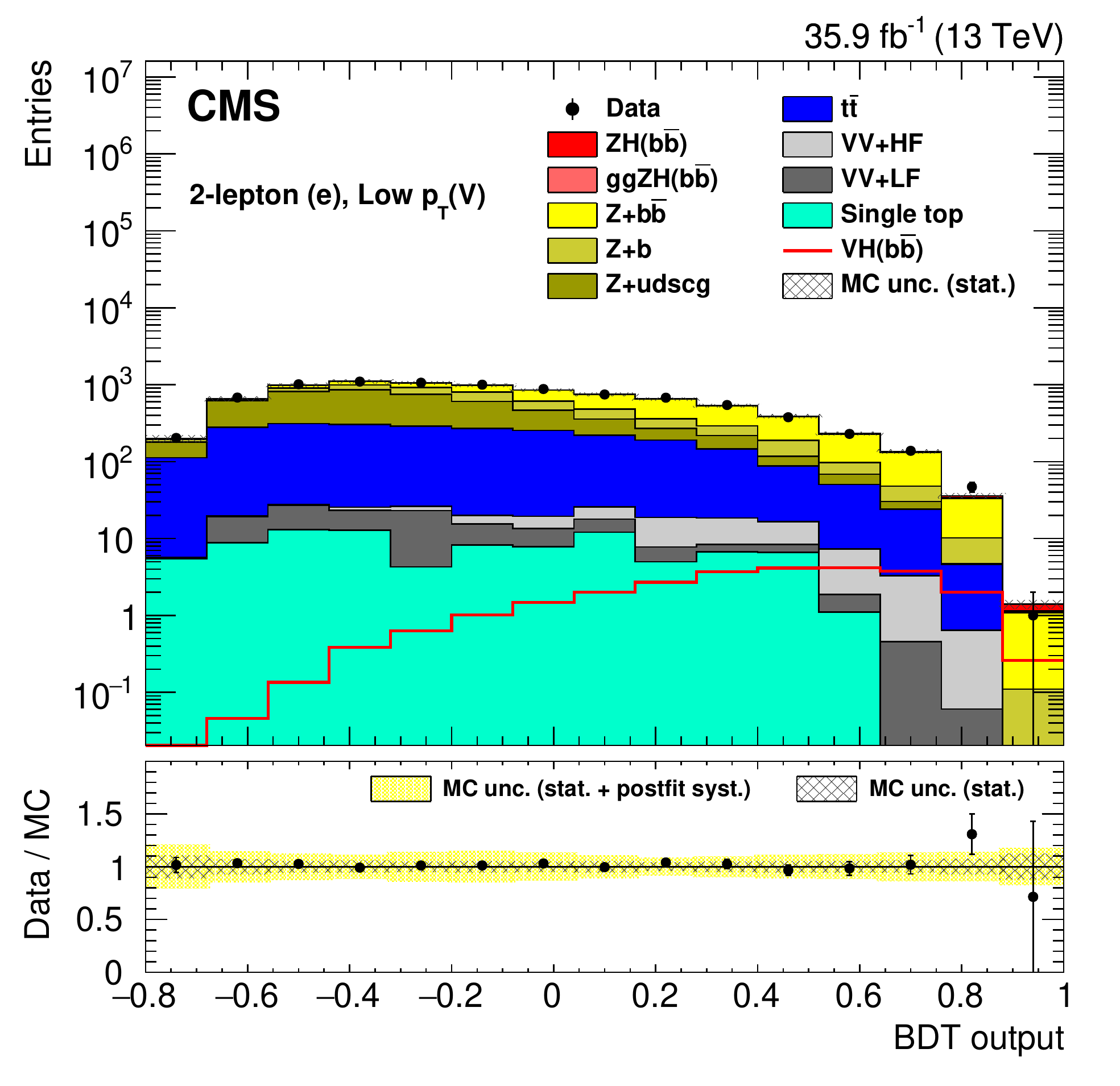}
    \includegraphics[width=0.33\textwidth]{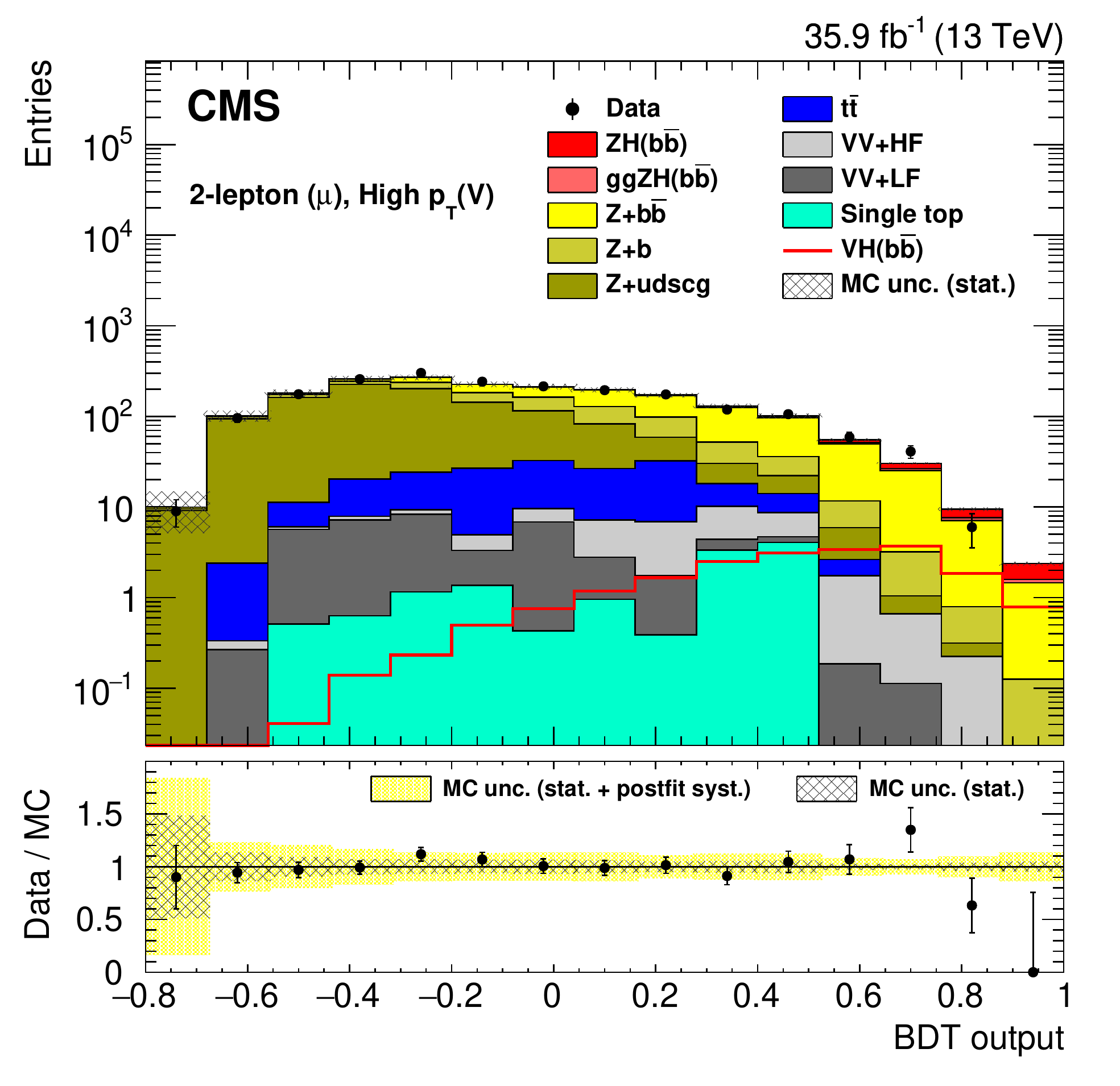}
    \includegraphics[width=0.33\textwidth]{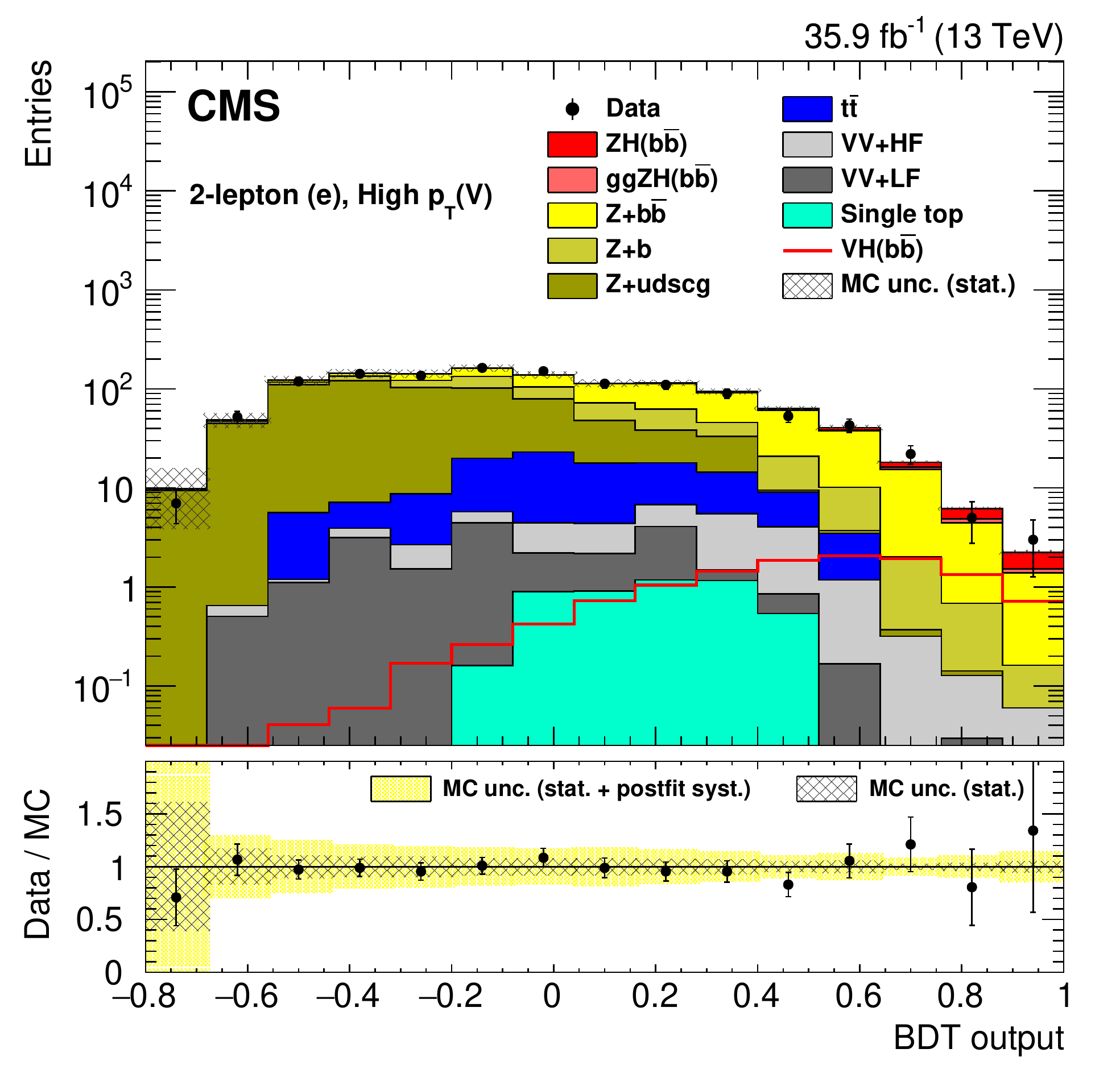}
      \caption{Post-fit event BDT output distributions for the 13\TeV data (points with error bars), for the \zerol\ channel (top), for the \onel\ channels (middle),
    and for the \twol\ low-\ptV  and high-\ptV regions (bottom).
The bottom inset shows the ratio of the number of events observed in data to that of the prediction from simulated samples for
the SM Higgs boson signal and for backgrounds.}
    \label{fig:post-fit-BDT}
\end{figure*}

\begin{figure}[htbp]
\centering
    \includegraphics[width=0.49\textwidth]{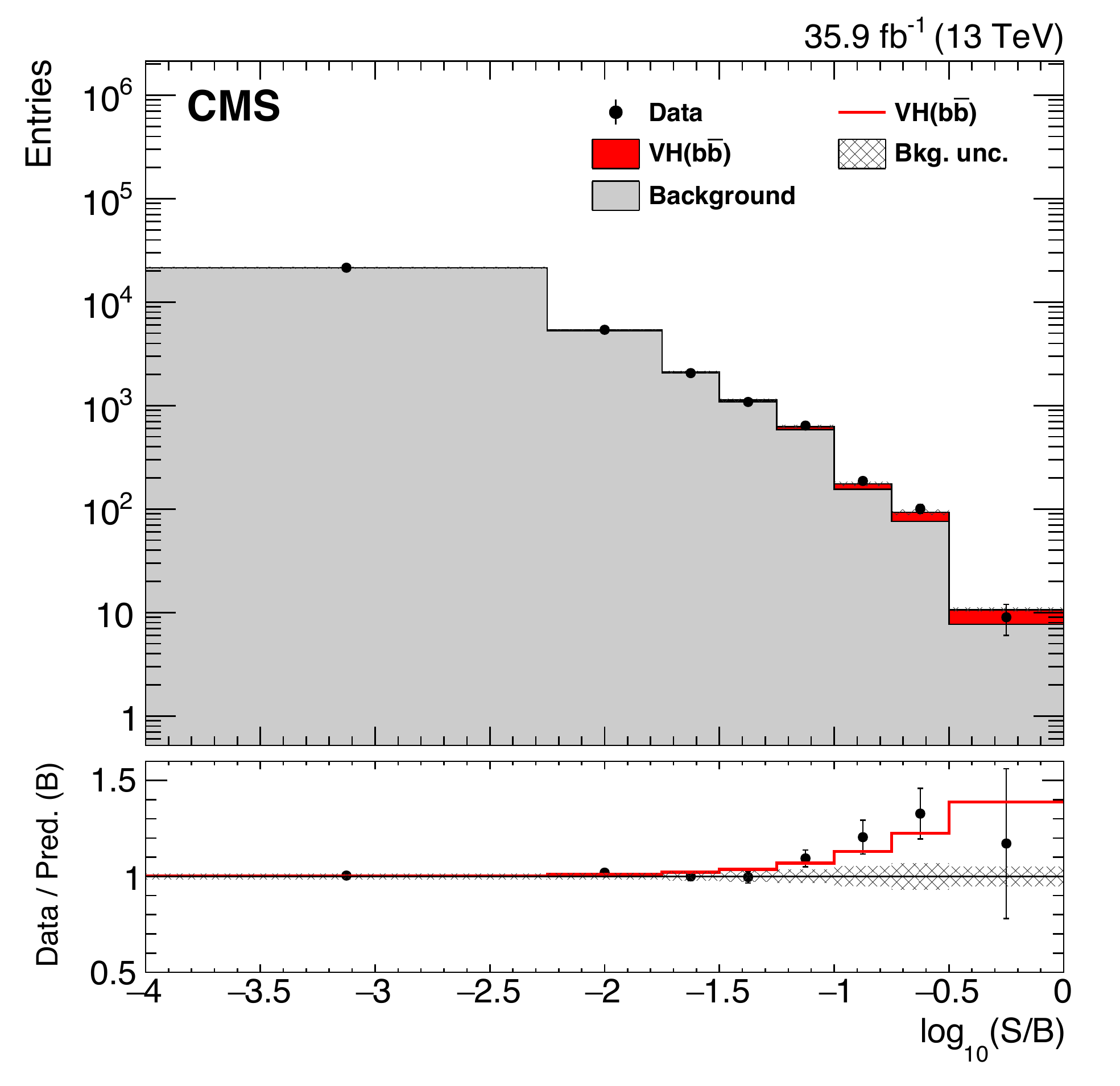}
\caption{Combination of all channels into a single event BDT distribution. Events are sorted in bins of similar expected
signal-to-background ratio, as given by the value of the output of their corresponding BDT
discriminant (trained with a Higgs boson mass hypothesis of 125\GeV).
The bottom plots show the ratio of the data to the background-only
prediction.}
    \label{fig:BDT_S_over_B_all}

\end{figure}

In total, seven event BDT output distributions are included in the fit:
one for the \zerol\ channel, one for each lepton flavor for the \onel\ channels, and two for each
lepton flavor for the \twol\ channels (corresponding to the two \ptV regions).  The number of
\CMVAmin\ distributions included is 24, corresponding to the control regions listed in
Tables~\ref{tab:ZnnControl}--\ref{tab:ZllControl}: three for the \zerol\ channel, four for each lepton flavor
for the \onel\ channels, and six for each lepton flavor for the \twol\ channels (each corresponding
to one of two \ptV regions).  Figure~\ref{fig:post-fit-BDT} shows the seven
BDT distributions after they have been adjusted by the fit. Figure~\ref{fig:BDT_S_over_B_all}
combines the BDT output values of all channels where the events
are gathered in bins of similar expected signal-to-background
ratio, as given by the value of the output of their corresponding BDT
discriminant. The
observed excess of events in the bins with the largest
signal-to-background ratio is consistent with what is expected from
the production of the SM Higgs boson. To detail this excess, the total numbers of events for all backgrounds, for the SM Higgs boson signal,
and for data are shown in Table~\ref{table:4bin_yields} for each channel, for the rightmost 20\% region of the BDT output distribution, where the sensitivity is large. The simulation yields are adjusted using the results of fit.

\begin{table*}
\topcaption{
The total numbers of events in each channel, for the rightmost 20\% region of the event BDT output distribution, are
shown for all background processes, for the SM Higgs boson \VH\ signal, and for data.  The yields from simulated samples
are computed with adjustments to the shapes and normalizations of the BDT distributions given by the signal extraction fit.
The signal-to-background ratio (S/B) is also shown.}
\centering
\begin{tabular}{lcccc} \hline
   Process       & \zerol\   & \onel\  & \twol\ low-\ptV  & \twol\ high-\ptV         \\
\hline
Vbb                &216.8 &102.5  &617.5    &113.9          \\
Vb                 &31.8  &20.0   &141.1    &17.2           \\
V+udscg            &10.2  &9.8    &58.4     &4.1            \\
\ttbar             &34.7  &98.0   &157.7    &3.2            \\
Single top quark   &11.8  &44.6   &2.3      &0.0            \\
VV(udscg)          &0.5   &1.5    &6.6      &0.5            \\
VZ(bb)         &9.9   &6.9    &22.9     &3.8            \\[\cmsTabSkip]
Total background   &315.7 &283.3  &1006.5   &142.7          \\
\VH                &38.3  &33.5   &33.7     &22.1           \\
Data               &334   &320    &1030     &179          \\[\cmsTabSkip]
S/B                &0.12  &0.12   &0.033    &0.15           \\
\hline
\end{tabular}\label{table:4bin_yields}
\end{table*}

The significance of the observed excess of events in the signal extraction fit is computed using
the standard LHC profile likelihood asymptotic approximation~\cite{junkcls,Read:2002hq,LHC-HCG-H-no-h,Cowan:2010js}. For $\mH =125.09$\GeV, it corresponds to a local significance of 3.3 standard deviations away from the background-only hypothesis.   This excess
is consistent with the SM prediction for Higgs boson production with signal
strength
$\mu = 1.19^{+0.21}_{-0.20}\stat^{+0.34}_{-0.32} \syst$.
The expected significance is 2.8 standard deviations with $\mu=1.0$. Together with this result, Table~\ref{tab:limits_by_mode} also lists the expected and observed significances for the \zerol\ channel, for the \onel\ channels combined, and for the \twol\ channels combined.

\begin{table}[htbp]
\topcaption{The expected and observed significances for \VH\ production with \HBB are shown, for $\mH=125.09$\GeV, for each channel fit
individually as well as for the combination of all three channels.
}
\label{tab:limits_by_mode}
\centering
\begin{tabular}{lcc}
\hline
Channels      & Significance  & Significance   \\
                    & expected      & observed       \\\hline
\zerol              &  1.5          &  0.0         \\
\onel               &  1.5          &  3.2          \\
\twol               &  1.8          &  3.1        \\[\cmsTabSkip]
Combined        &  2.8          &  3.3      \\\hline
\end{tabular}
\end{table}

\begin{figure}[htbp]
  \centering
\includegraphics[width=\cmsFigWidth]{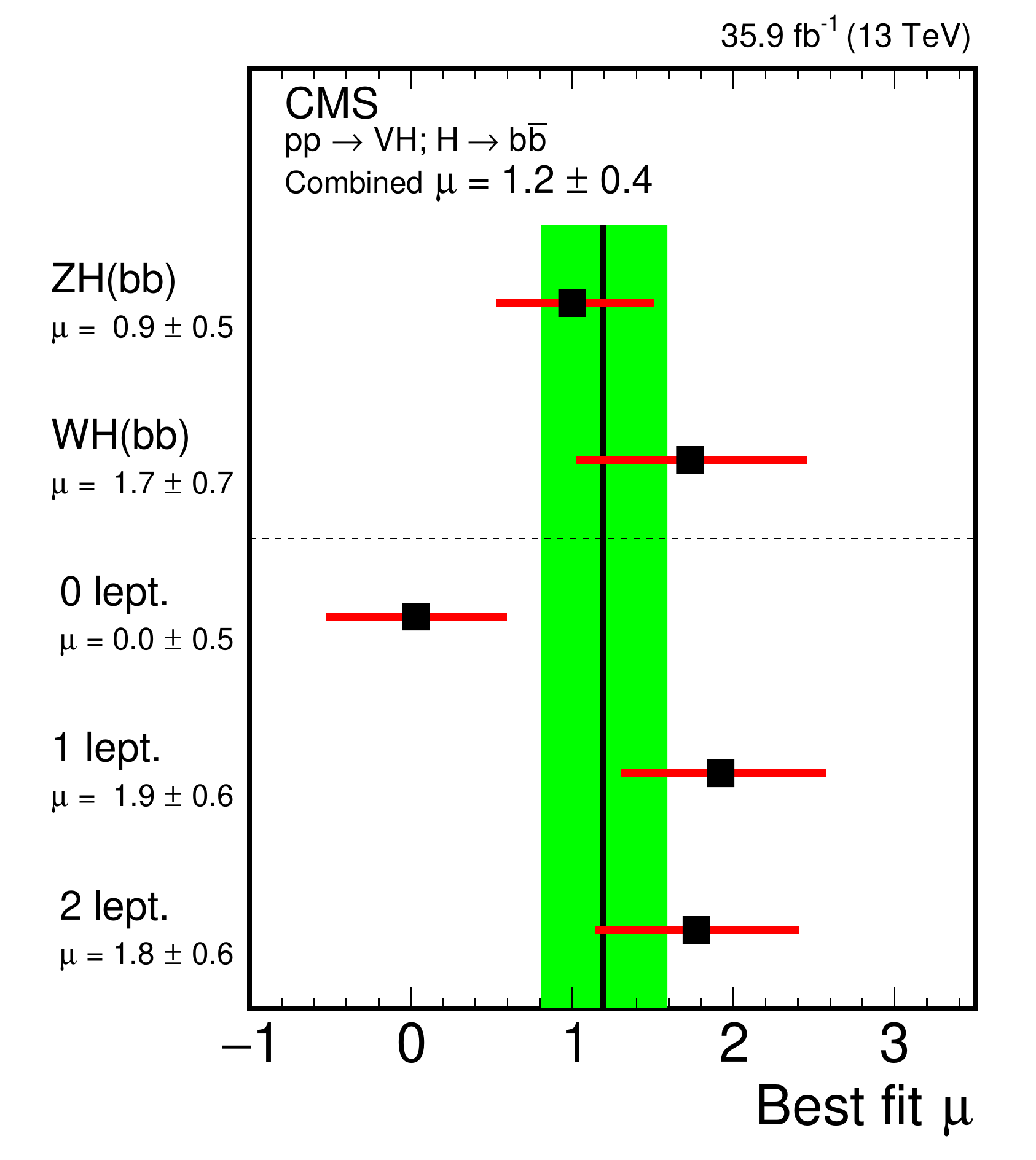}
\caption{ The best fit value of the signal
strength $\mu$, at $\mH=125.09$\GeV, is shown in black with a green uncertainty band.  Also shown are the results of a separate fit where each channel is assigned an independent signal strength parameter. Above the dashed line
are the \WH\ and \ZH\ signal strengths derived from a fit where each production mode is assigned an independent signal strength parameter.}
    \label{fig:mu-values}

\end{figure}

The observed signal strength $\mu$ is shown in the lower portion of Fig.~\ref{fig:mu-values}
for \zero, \one\ and \twol\ channels.  The observed signal strengths of the three channels
are consistent with the combined best fit signal strength with a probability of 5\%.
In the upper portion of Fig.~\ref{fig:mu-values}
the signal strengths for the separate \WH\ and \ZH\ production processes are
shown.  The two production modes are consistent with the SM
expectations within uncertainties. The fit for the \WH\ and \ZH\ production modes is not fully correlated to the analysis channels
because the analysis channels contain mixed processes.  The \WH\ process contributes
approximately 15\%  of the Higgs boson signal event yields in the \zerol\ channel,
resulting from events in which the lepton is outside the detector acceptance, and the
ZH process contributes less than 3\% to the \onel\ channel when one of the
leptons is outside the detector acceptance.

\begin{figure}[htbp]
   \centering
    \includegraphics[width=0.48\textwidth]{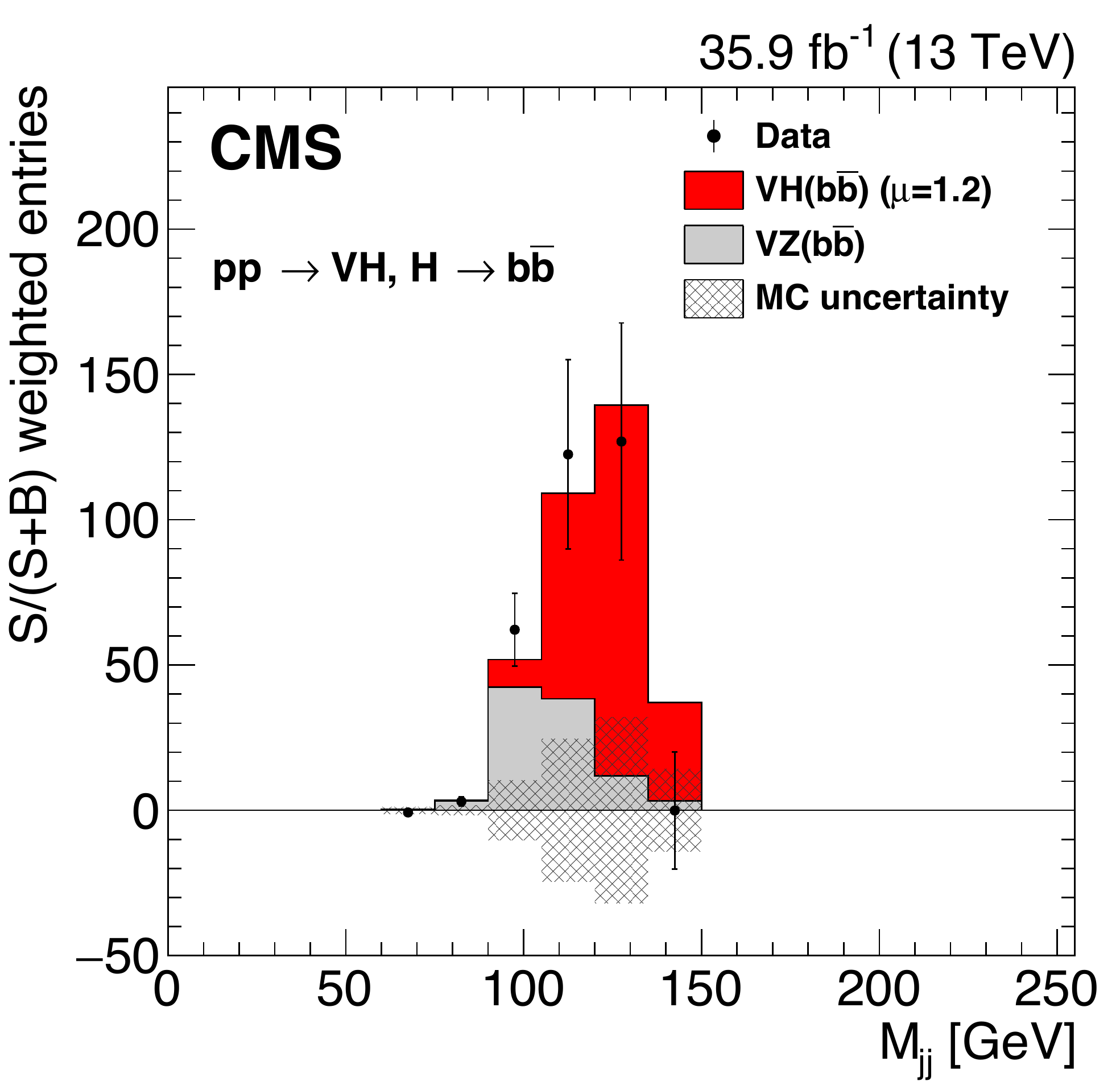}
 \caption{Weighted dijet invariant mass distribution for events in all channels combined.
    Shown are data and the \VH\ and \VZ\ processes with all other background processes
    subtracted. Weights are derived from the event BDT output distribution as described in the text.
 }
     \label{fig:mjj_all}

\end{figure}

Figure~\ref{fig:mjj_all} shows a dijet invariant mass distribution, combined for all channels, for data and for the \VH\ and \VZ\ processes, with all other background processes subtracted. The distribution is constructed from all events that populate the signal region event BDT distributions shown in Fig.~\ref{fig:post-fit-BDT}. The values of the scale factors and nuisance parameters from the fit used to extract the \VH\ signal are propagated to this distribution. To better visualize the contribution of events from signal, all events are weighted by $\mathrm{S/(S{+}B)}$, where S and B are the numbers of expected signal and total post-fit background events in the bin of the output of the BDT distribution in which each event is contained. The data are consistent with
the production of a standard model Higgs boson decaying to \bbbar. In the Figure, aside from the weights, which favor the \VH\ process, the event yield from \VZ\ processes is reduced significantly due to the \ptV and \Mjj\ selection requirements for the \VH\ signal region, and from the training of the BDT that further discriminates against diboson processes.

\subsection{Extraction of \texorpdfstring{\VZ}{} with \texorpdfstring{\ZBB}{}}\label{sec:diboson}

The \VZ\ process with \ZBB, having a nearly identical final state as \VH\ with \HBB, serves as a validation of the methodology used in the search for the latter process.
To extract this diboson signal, event BDT discriminants are trained using as signal the
simulated samples for this process. All other processes, including
\VH\ production (at the predicted SM rate), are treated as background. The only modification made is the requirement that
the signal region \Mjj\ be in the $[60,160]$\GeV range.

The results from the combined fit for all channels of the control and signal region distributions, as defined
in Sections~\ref{sec:hbb_SR} and~\ref{sec:hbb_CR},
are summarized in Table~\ref{tab:vvresults}
for the same $\sqrt{s}=13\TeV$ data used in the \VH\ search described above. The observed excess of events for the combined $\PW\cPZ$ and $\cPZ\cPZ$ processes has a significance of 5.0 standard deviations from the background-only
event yield expectation. The corresponding signal
strength, relative to the prediction of the {\MGvATNLO} generator at NLO
mentioned in
Section~\ref{sec:hbb_Simulations},
is measured to be  $\mu_{\Vvar\Vvar} = {1.02}_{-0.23}^{+0.22}$.

\begin{table*}[htbp]
\topcaption{Validation results for \VZ\ production with \ZBB.  Expected and observed significances, and the observed signal strengths. Significance values are given in numbers of standard deviations.}
\label{tab:vvresults}
\centering
{
\begin{tabular}{lccc}
\hline
Channels   & Significance  & Significance  & Signal strength  \\
          & expected      & observed      & observed         \\
\hline
\zerol    & 3.1           & 2.0           & $0.57\pm0.32$    \\
\onel     & 2.6           & 3.7           & $1.67\pm0.47$    \\
\twol     & 3.2           & 4.5           & $1.33\pm0.34$    \\[\cmsTabSkip]
Combined  & 4.9           & 5.0           & $1.02\pm0.22$    \\
\hline
\end{tabular}
}
\end{table*}

\begin{figure}[htbp]
  \centering
    \includegraphics[width=0.49\textwidth]{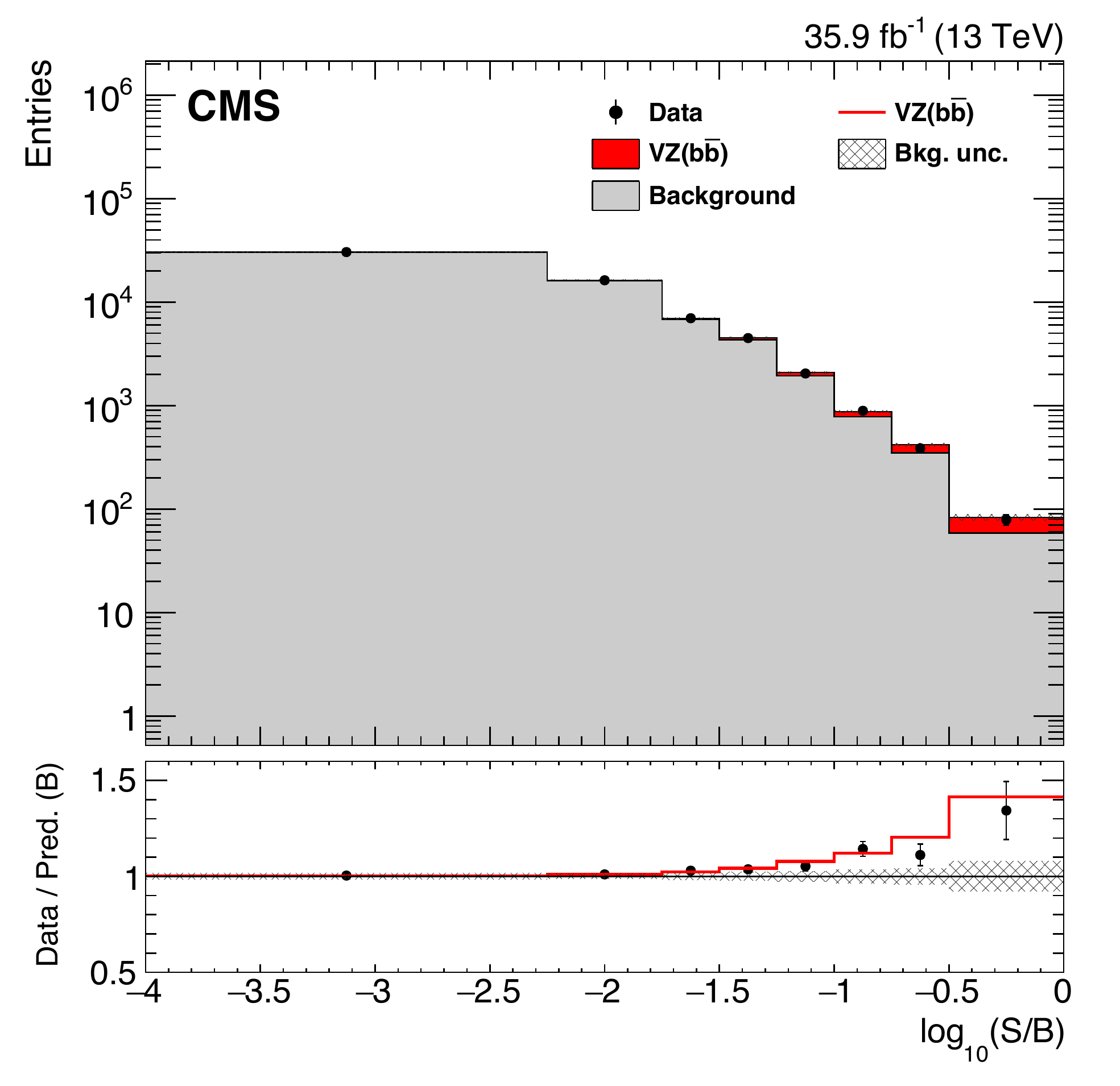}
    \caption{
Combination of all channels in the \VZ\ search, with \ZBB\, into a single event BDT distribution. Events are sorted in bins of similar expected signal-to-background ratio, as given by the value of the output of their corresponding BDT discriminant. The bottom inset shows the ratio of the data to the predicted background, with a red line overlaying the expected SM contribution from \VZ\ with \ZBB.}
    \label{fig:logSB_VV}

\end{figure}

Figure~\ref{fig:logSB_VV} shows the combined event BDT output distribution for all channels,
with the content of each bin, for each channel, weighted by the expected signal-to-background
ratio.  The excess of events in data, over background, is shown to be compatible with the
yield expectation from \VZ\ production with \ZBB.

\subsection{Combination with Run~1 VH(bb) analysis }

The results from the search for \VH\ with \HBB, presented in this article, are combined
with those from the similar searches performed by the CMS experiment~\cite{Chatrchyan:1610290,Khachatryan:2015bnx,Khachatryan:2016vau} during Run~1 of the LHC, using proton-proton collisions at $\sqrt{s}=7$ and
8\TeV with data samples corresponding to integrated luminosities of up to 5.1
and 18.9\fbinv, respectively. The combination yields an observed  signal significance, at $\mH =125.09$\GeV, of 3.8 standard
deviations, where 3.8 are expected. The corresponding signal strength is
$\mu=1.06^{+0.31}_{-0.29}$. All systematic
uncertainties are assumed to be uncorrelated in the combination, except for cross section uncertainties
derived from theory, which are assumed to be fully correlated. Treating all
uncertainties as uncorrelated has a negligible effect on the significance. Table~\ref{tab:results_combo} lists these results.

\begin{table*}
\topcaption{The expected and observed significances and the observed signal strengths
for \VH\ production with \HBB for Run~1 data~\cite{Khachatryan:2016vau}, Run~2 (2016) data,
and for the combination of the two.  Significance values are given in numbers of standard deviations.
}
\label{tab:results_combo}
\centering
\begin{tabular}{lccc}
\hline
Data used       & Significance  & Significance & Signal strength  \\
                    & expected      & observed     &  observed\\\hline\\[-2.2ex]
Run~1                &  2.5          &  2.1         &  $0.89_{-0.42}^{+0.44}$ \\[0.6ex]
Run~2                &  2.8          &  3.3         &  $1.19_{-0.38}^{+0.40}$ \\[0.6ex]
Combined             &  3.8          &  3.8         &  $1.06_{-0.29}^{+0.31}$ \\[0.4ex]
\hline
\end{tabular}
\end{table*}

\section{Summary}\label{sec:hbb_Conclusions}

A search for the standard model (SM) Higgs boson (H) when produced in association with an electroweak
vector boson and decaying to a \bbbar pair is reported for the
$\cPZ(\nu\nu)\PH$, $\PW(\mu\nu)\PH$, $\PW(\Pe\nu)\PH$,
$\cPZ(\mu\mu)\PH$, and $\cPZ(\Pe\Pe)\PH$ processes.
The search is performed in data samples corresponding to an integrated
luminosity of
35.9\fbinv at $\sqrt{s}=13\TeV$, recorded by the CMS experiment at the LHC.
 The observed signal significance, for $\mH=125.09$\GeV, is 3.3 standard deviations, where the expectation from the
 SM Higgs boson production is 2.8. The corresponding signal strength is $\mu=1.2\pm0.4$.

The combination of this
result with the one from the same measurement performed by the CMS Collaboration in Run~1 of the LHC
using proton-proton collisions at $\sqrt{s}=7$ and 8\TeV with data samples
corresponding to integrated luminosities of up to 5.1 and 18.9\fbinv, respectively, yields an
observed signal significance of 3.8 standard deviations, where 3.8 are expected from the SM
signal. The corresponding signal strength is
$\mu=1.06^{+0.31}_{-0.29}$.

The result presented in this article provides evidence for the decay of the Higgs boson into a pair of b quarks with a rate consistent with the SM expectation.

\begin{acknowledgments}
We congratulate our colleagues in the CERN accelerator departments for the excellent performance of the LHC and thank the technical and administrative staffs at CERN and at other CMS institutes for their contributions to the success of the CMS effort. In addition, we gratefully acknowledge the computing centers and personnel of the Worldwide LHC Computing Grid for delivering so effectively the computing infrastructure essential to our analyses. Finally, we acknowledge the enduring support for the construction and operation of the LHC and the CMS detector provided by the following funding agencies: BMWFW and FWF (Austria); FNRS and FWO (Belgium); CNPq, CAPES, FAPERJ, and FAPESP (Brazil); MES (Bulgaria); CERN; CAS, MoST, and NSFC (China); COLCIENCIAS (Colombia); MSES and CSF (Croatia); RPF (Cyprus); SENESCYT (Ecuador); MoER, ERC IUT, and ERDF (Estonia); Academy of Finland, MEC, and HIP (Finland); CEA and CNRS/IN2P3 (France); BMBF, DFG, and HGF (Germany); GSRT (Greece); OTKA and NIH (Hungary); DAE and DST (India); IPM (Iran); SFI (Ireland); INFN (Italy); MSIP and NRF (Republic of Korea); LAS (Lithuania); MOE and UM (Malaysia); BUAP, CINVESTAV, CONACYT, LNS, SEP, and UASLP-FAI (Mexico); MBIE (New Zealand); PAEC (Pakistan); MSHE and NSC (Poland); FCT (Portugal); JINR (Dubna); MON, RosAtom, RAS, RFBR and RAEP (Russia); MESTD (Serbia); SEIDI, CPAN, PCTI and FEDER (Spain); Swiss Funding Agencies (Switzerland); MST (Taipei); ThEPCenter, IPST, STAR, and NSTDA (Thailand); TUBITAK and TAEK (Turkey); NASU and SFFR (Ukraine); STFC (United Kingdom); DOE and NSF (USA).

\hyphenation{Rachada-pisek} Individuals have received support from the Marie-Curie program and the European Research Council and Horizon 2020 Grant, contract No. 675440 (European Union); the Leventis Foundation; the A. P. Sloan Foundation; the Alexander von Humboldt Foundation; the Belgian Federal Science Policy Office; the Fonds pour la Formation \`a la Recherche dans l'Industrie et dans l'Agriculture (FRIA-Belgium); the Agentschap voor Innovatie door Wetenschap en Technologie (IWT-Belgium); the Ministry of Education, Youth and Sports (MEYS) of the Czech Republic; the Council of Science and Industrial Research, India; the HOMING PLUS program of the Foundation for Polish Science, cofinanced from European Union, Regional Development Fund, the Mobility Plus program of the Ministry of Science and Higher Education, the National Science Center (Poland), contracts Harmonia 2014/14/M/ST2/00428, Opus 2014/13/B/ST2/02543, 2014/15/B/ST2/03998, and 2015/19/B/ST2/02861, Sonata-bis 2012/07/E/ST2/01406; the National Priorities Research Program by Qatar National Research Fund; the Programa Severo Ochoa del Principado de Asturias; the Thalis and Aristeia programs cofinanced by EU-ESF and the Greek NSRF; the Rachadapisek Sompot Fund for Postdoctoral Fellowship, Chulalongkorn University and the Chulalongkorn Academic into Its 2nd Century Project Advancement Project (Thailand); the Welch Foundation, contract C-1845; and the Weston Havens Foundation (USA). \end{acknowledgments}

\bibliography{auto_generated}

\cleardoublepage \appendix\section{The CMS Collaboration \label{app:collab}}\begin{sloppypar}\hyphenpenalty=5000\widowpenalty=500\clubpenalty=5000\textbf{Yerevan Physics Institute,  Yerevan,  Armenia}\\*[0pt]
A.M.~Sirunyan, A.~Tumasyan
\vskip\cmsinstskip
\textbf{Institut f\"{u}r Hochenergiephysik,  Wien,  Austria}\\*[0pt]
W.~Adam, F.~Ambrogi, E.~Asilar, T.~Bergauer, J.~Brandstetter, E.~Brondolin, M.~Dragicevic, J.~Er\"{o}, A.~Escalante Del Valle, M.~Flechl, M.~Friedl, R.~Fr\"{u}hwirth\cmsAuthorMark{1}, V.M.~Ghete, J.~Grossmann, J.~Hrubec, M.~Jeitler\cmsAuthorMark{1}, A.~K\"{o}nig, N.~Krammer, I.~Kr\"{a}tschmer, D.~Liko, T.~Madlener, I.~Mikulec, E.~Pree, N.~Rad, H.~Rohringer, J.~Schieck\cmsAuthorMark{1}, R.~Sch\"{o}fbeck, M.~Spanring, D.~Spitzbart, W.~Waltenberger, J.~Wittmann, C.-E.~Wulz\cmsAuthorMark{1}, M.~Zarucki
% ...
\vskip\cmsinstskip
% ...
1:~~Also at Vienna University of Technology, Vienna, Austria\\
% ...

\end{sloppypar}
\end{document}